\pdfoutput=1
\documentclass{llncs}
\usepackage{graphicx}
\usepackage[boxruled]{algorithm2e}
\usepackage{amsmath}
\usepackage{amssymb}
\usepackage{wrapfig}
\usepackage{array}
\usepackage{hhline}
\usepackage{xspace}
\usepackage{ragged2e}
\usepackage{url}
\newcommand{\measure}[1]{\ensuremath{\text{\normalfont{\sffamily {#1}}}}\xspace}
\newcommand{\C}{\ensuremath{\mathcal{C}}}
\newcommand{\moveVC}[2]{\ensuremath{\C_{#1 \to #2}}}
\newcommand{\w}{m}
\newcommand{\algo}[1]{\textsc{#1}}

\newcommand{\andreapar}{\vspace*{.5ex}\par\noindent}

\SetAlgoCaptionLayout{robscaptionstyle}
\SetAlgorithmName{Algo.}{algorithmautorefname}{loa}
\usepackage{float}
\newfloat{myalgo}{tbhp}{mya}

\usepackage{subfig}
\graphicspath{{./pics/}}
\title{Experiments on Density-Constrained Graph Clustering\thanks{This work was partially supported by the DFG under grant WA 654/19-1}}
\author{Robert G\"orke \and Andrea Schumm \and Dorothea Wagner}
\institute{Institute of Theoretical Informatics, Karlsruhe Institute of Technology, Germany}
\begin{document}
\maketitle
\setcounter{totalnumber}{8}
\setcounter{topnumber}{8}
\setcounter{bottomnumber}{8}
\renewcommand{\textfraction}{0.0}
\renewcommand{\topfraction}{1.0}
\renewcommand{\bottomfraction}{1.0}
\begin{abstract}
Clustering a graph means identifying internally dense subgraphs which are only sparsely interconnected.
Formalizations of this notion lead to measures that quantify the quality of a clustering and to algorithms that actually find clusterings.
Since, most generally, corresponding optimization problems are hard, heuristic clustering algorithms are used in practice, or other approaches which are not based on an objective function.
In this work we conduct a comprehensive experimental evaluation of the qualitative behavior of greedy bottom-up heuristics driven by cut-based objectives and constrained by intracluster density, using both real-world data and artificial instances.
Our study documents that a greedy strategy based on local movement is superior to one based on merging.
We further reveal that the former approach generally outperforms alternative setups and reference algorithms from the literature in terms of its own objective, while a modularity-based algorithm competes surprisingly well.
Finally, we exhibit which combinations of cut-based inter- and intracluster measures are suitable for identifying a hidden reference clustering in synthetic random graphs.
\end{abstract}

\section{Introduction}\label{sec:intro}
Graph clustering aims at finding subsets of vertices that are densely connected with each other but sparsely connected with the remainder of the graph.
In the last decades, interest in graph clustering algorithms has grown rapidly, with applications ranging from customer recommendation systems to the analysis of networks describing social ties or protein-protein interaction. 
A variety of measures have been proposed, which are used to assess and compare different clusterings and to guide the design of algorithms.
Traditional methods from algorithmics often focus on sparse cuts with respect to measures like conductance \cite{kvv-cgds-00} or expansion \cite{hlw-e-06}, while, 
independent from that, a measure called modularity \cite{ng-fecsn-04} proved to yield meaningful clusterings on a wide range of application data. 

Recently, we systematically assembled a range of self-evident intracluster density and intercluster sparsity measures for clusterings, where the latter are based on conductance , expansion  and density of the cuts induced by the clusters \cite{gsw-dcgc-11b}.
We further formally stated the problem \textsc{Density-Constrained Clustering} (DCC), where the objective is to optimize intercluster sparsity with the constraint that the intracluster density must exceed a given threshold.   
As optimal polynomial-time algorithms for \textsc{DCC} are unknown, we investigated how different combinations of intracluster sparsity and intercluster density measure influence the efficiency of a greedy optimization strategy based on cluster merging. 
However, little is known about its qualitative behavior in practical scenarios, and an experimental evaluation of \textsc{DCC} has yet been missing.
\andreapar\textbf{Our Contribution.}
We provide a comprehensive study of the practical behavior of greedy graph clustering heuristics driven by cut-based objectives and constrained by intracluster density. 
We give evidence that, in general, greedy algorithms based on local vertex moves lead to better quality than the corresponding merge-based algorithm. 
We then compare the move-based algorithm to a set of reference algorithms from the literature, both with respect to the objective of \textsc{DCC} and their ability to reconstruct planted partitions in a family of synthetic graphs. 
We find that the greedy move algorithm compares favorably to most reference algorithms in the context of \textsc{DCC}, while a comparison with the modularity-based algorithm shows that optimizing modularity implicitly yields good results for some variants of \textsc{DCC}.
Experiments with planted partition graphs suggest that certain combinations of inter- and intracluster measures are effective in finding the hidden clustering, while others clearly fail.
Together with observations about the number of identified clusters, this yields valuable insights about the behavior of the respective intra- and intercluster density measures.
\andreapar\textbf{Related Work.}
Related clustering algorithms are Iterative Conductance Cutting~\cite{kvv-cgds-00}, Markov-Clustering \cite{phd-dongen-02}, Geometric MST Clustering~\cite{bgw-egca-03} and a modularity-based greedy algorithm based on vertex moves~\cite{rn-m-11};
we use these as reference algorithms.
Kannan et al.\ propose to minimize the cut between, subject to a guaranteed conductance within clusters \cite{kvv-cgds-00}, which is closely related to the DCC. 
They further show that Iterative Conductance Cutting has polylogarithmic approximation guarantees on both of these measures.
Brandes et al.\ conduct an experimental study on the performance of Iterative Conductance Cutting, Markov-Clustering and Geometric MST Clustering, both with respect to quality and running times \cite{bgw-egcme-07}.
A similar, but more recent study can be found in \cite{lf-ca-09}.
Flake et al.\ give a clustering algorithm with provable, but interdependent bounds on both intra- and a variant of intercluster expansion.
The notion of modularity was introduced in \cite{ng-fecsn-04}, an extensive and recent overview of the research on it can be found in \cite{f-c-09}.
Apart from these, there is a huge number of publications on graph clustering, for an overview see \cite{jd-acd-88,b-ascdm-06}.    
\section{Preliminaries}

\textbf{Notation.}
Let $G=(V,E)$ be an undirected, unweighted, and simple graph, i.e.~$G$ is loopless and has not parallel edges. 
In the following, $n$ will always denote the number of vertices and $m$ the number of edges in $G$.
For two subsets $A$ and $B$ of $V$, $\w_{A,B} := |\bigl\{\{u, v\} \in E \mid u \in A, v \in B\bigr\}|$ is the number of edges between $A$ and $B$, 
$n_A := |A|$ is the number of vertices in $A$, $\w_A :=|E(A)|$ is its number of intracluster edges and $x_A := \w_{A,V \setminus A}$ the number of intercluster edges incident to $A$.  
Further, the \emph{volume}  $v_A$ of $A$ is defined as $v_A := \sum_{v \in A} \deg(v)$. 
\setlength{\tabcolsep}{4pt}
\begin{wraptable}[26]{r}{.5\textwidth}
\small
\vspace*{-5.5ex}
\caption{Density measures}
\vspace*{-2ex}
\label{tab:measures:1}
\begin{tabular}{|l|l|l|}
\hline

\multicolumn{3}{|c|}{\rule{0pt}{2.4ex}intracluster density}\\

\hline
\rule{0pt}{3ex}global & \measure{gid} &$\frac{\sum_{C \in \C} m_C}{\sum_{C \in \C} \binom{n_C}{2}}$\\
\rule{0pt}{3ex}minimum & \measure{mid} & $\min\limits_{C \in \C}\frac{m_C}{\binom{n_C}{2}}$\\
\rule{0pt}{3ex}average & \measure{aid} & $\frac{1}{|\C|} \sum\limits_{C \in \C} \frac{m_C}{\binom{n_C}{2}}$\\
\hline
\hline

\multicolumn{3}{|c|}{\rule{0pt}{2.4ex}intercluster density} \\

\hline
\rule{0pt}{3ex}global & \measure{gxd} &$\frac{\sum_{A \neq B \in \C} m_{A, B}}{\sum_{A \neq B \in \C} n_A n_B}$\\
\rule{0pt}{3ex}maximum & \measure{mixd} & $\max\limits_{C \in \C} \frac{x_{C}}{n_C n_{V \setminus C}}$\\
\rule{0pt}{2.5ex}average& \measure{aixd} & $\frac{1}{|\C|} \!\! \sum\limits_{C \in \C}\!\! \frac{x_{C}}{n_C  n_{V \setminus C}}$\\
\hline
\hline
\multicolumn{3}{|c|}{\rule{0pt}{2.4ex}intercluster conductance} \\

\hline
\rule{0pt}{3ex}maximum & \measure{mixc} & $\max\limits_{C \in \C} \frac{m_{C, V\setminus C}}{\min\{v_C, v_{V \setminus C}\}}$\\
\rule{0pt}{2.5ex}average & \measure{aixc} & $\frac{1}{|\C|}\sum\limits_{C \in \C} \frac{m_{C, V\setminus C}}{\min\{v_C, v_{V \setminus C}\}}$\\
\hline
\hline
\multicolumn{3}{|c|}{\rule{0pt}{2.4ex}intercluster expansion} \\

\hline
\rule{0pt}{3ex}maximum & \measure{mixe} & $\max\limits_{C \in \C} \frac{m_{C, V\setminus C}}{\min\{n_C, n_{V \setminus C}\}}$\\
\rule{0pt}{2.5ex}average& \measure{aixe} & $\frac{1}{|\C|}\sum\limits_{C \in \C} \frac{m_{C, V\setminus C}}{\min\{n_C, n_{V \setminus C}\}}$\\
\hline
\hline
 \multicolumn{3}{|c|}{\rule{0pt}{2.4ex}intercluster edges} \\

\hline
global &  \measure{nxe} & $\sum_{A \neq B \in \C} m_{A, B}$  \\
\hline 
\hline
 \multicolumn{3}{|c|}{\rule{0pt}{2.4ex}modularity} \\

\hline
global &  \measure{mod} & $\frac{ \sum_{C \in \C} m_C}{m} -  \frac{ \sum_{C \in \C}  v_C^2}{4m^2}$  
\\
\hline 
\end{tabular}
\normalsize
\end{wraptable}
The \emph{conductance} of a cut $(S,T)$ measures the bottleneck between $S$ and $T$, defined as $\frac{\w_{S,T}}{\min \{v_S, v_T\}}$;
\emph{expansion} substitutes volume by cardinality: $\frac{\w_{S,T}}{\min \{n_S, n_T\}}$.
The \emph{density} (or \emph{sparsity}) of a cut is $\frac{\w_{S,T}}{n_S  n_T}$, which equals the \emph{uniform minimum-ratio cut}.
We restrict ourselves to disjoint clusters in this work, this means, if $\mathcal{C} = \{C_1 , \dots , C_k\}$ is a partition of~$V$, we call~$\mathcal{C}$ a \emph{clustering} of $G$ and the sets~$C_i$ \emph{clusters}.
The cluster containing vertex $v$ is $\C(v)$ and the clustering that results from moving vertex $v$ to cluster $D$, i.e.~$\bigl(\C \setminus \{\C(v), D\}\bigr) \cup \{\C(v) \setminus v, D \cup \{v\}\}$, is abbreviated by $\moveVC{v}{D}$.
A clustering is \emph{trivial} if either~$k=1$ (\emph{all-clustering}), or each cluster contains only one element (\emph{singletons}).
We identify a cluster~$C$ with the set of nodes it constitutes and with its vertex-induced subgraph of $G$.
Then $E({\cal C}):=\bigcup_{C \in \C} E(C)$ are called \emph{intracluster} edges and~$E\setminus {E(\mathcal{C})}$ \emph{intercluster} edges.
A \emph{clustering measure} is a function that maps clusterings to real numbers, thereby assessing the quality of a clustering.
We define high quality to correspond to high (low) values of intracluster (intercluster) measures and will always denote intracluster density measures with $i$ and intercluster density measures with $x$, unless otherwise stated.

\andreapar\textbf{Intracluster Density and Intercluster Sparsity Measures.}
All intercluster measures we use are based on \emph{cuts} or \emph{$k$-way cuts}. 
Separating a single cluster from the remaining vertices induces a cut, 
whose sparsity can be evaluated using density, conductance or expansion.
This defines a set of sparsity values for the whole clustering, from which we can either compute the average or the maximum, yielding \emph{maximum/average intercluster density/conductance/expansion} (\measure{mixd}, \measure{aixd}, \measure{mixc}, \measure{aixc}, \measure{mixe} and \measure{aixe})\footnote{Note that we keep the $i$ in the abbreviations, although in contrast to \cite{gsw-dcgc-11b}, we do not distinguish between pairwise and isolated measures}.
Another point of view is to evaluate the clustering as a whole, i.e.~to assess the sparsity of the induced $k$-way cut directly.
We do this by either counting the number of intercluster edges (\measure{nxe}) or by dividing the number of intercluster edges by the maximum possible number, i.e.~the number of intercluster pairs (\measure{gxd}).
It is possible to use similar, cut-based measures for intracluster density. 
However, even evaluating these measures for a given clustering is $NP$-hard, such that clustering algorithms usually work with approximations or bounds \cite{kvv-cgds-00,ftt-gcmct-04,bgw-egcme-07}.
As we intend to use intracluster density measures as constraints in greedy bottom-up algorithms, it is crucial to be able to evaluate them efficiently. 
We therefore use a more practical approach and define intracluster density as the ratio of the number of intracluster edges and the number of intracluster pairs. 
Evaluating this globally leads to \emph{global intracluster density} (\measure{gid}), whereas the average and minimum of all clusters yields \emph{average} and \emph{minimum intracluster density} (\measure{aid} and \measure{mid}).
\par
Table \ref{tab:measures:1} summarizes the formalizations of all 
measures considered.
Note that, in contrast to the set of measures used in \cite{gsw-dcgc-11b}, we omit the notions of \emph{pairwise} densities as they turned out to be very prone to local minima if used with greedy bottom-up algorithms.
Although it does not quite fit into this classification, Table \ref{tab:measures:1} also includes the objective used by one of the reference algorithms, \emph{modularity}, which simultaneously assesses intracluster density and intercluster sparsity by subtracting from the fraction of intracluster edges the expectation of this value in a random graph (high modularity corresponds to high quality).
\setlength{\extrarowheight}{0pt}
\andreapar\textbf{Density-Constrained Clustering.}
Density-Constrained Clustering is the problem of optimizing intercluster density while retaining guarantees on the intracluster density.
Considering each combination of intracluster and intercluster measure listed in Table \ref{tab:measures:1} leads to a family of optimization problems.
Slightly abusing the notation, we consider modularity as an intercluster density objective in this context.
\vspace{-2ex}
\begin{problem}[\algo{Density-Constrained Clustering(DCC)}]
  Given a graph $G=(V, E)$, among all clusterings with an intracluster density of no less than $\alpha$, find a clustering $\C$ with optimum intercluster quality.
\end{problem}
\section{Greedy Algorithms for \textsc{Density-Constrained Clustering}}
The following generic greedy algorithms heuristically minimize(maximize) the objective function of \textsc{DCC} for all density measures considered.  
\andreapar\textbf{Greedy Merge (GM).}
Starting from singletons, the algorithm greedily merges pairs of clusters. In each step, among all pairs of clusters whose merge does not violate the constraint on the intracluster density, the merge with the largest benefit to the intercluster density is performed.
We recently proposed this algorithm in the context of DCC~\cite{gsw-dcgc-11b} and  classified combinations of intercluster and intracluster density with respect to the question how efficiently this algorithm can be implemented.
Algorithms of these kind are common in the context of clustering point sets in $d$-dimensional space, where a basic constraint is that the number of clusters must not fall below a certain threshold. In the field of graph clustering, this algorithm is used to optimize modularity \cite{cnm-fcsln-04}.
\begin{myalgo}%
\noindent\begin{minipage}[b]{0.48\textwidth}
\begin{algorithm}[H]\footnotesize
 \caption{\textsc{Greedy Vertex Moving}}
 \label{algo:GLM}
 \SetKwInOut{Input}{Input}\SetKwInOut{Output}{Output}
 \Input{graph $G$, \measure{inter}, \measure{intra}, $\alpha$}
 \Output{clustering $\C_0$ of $G$}
\DontPrintSemicolon
  $G^0 \leftarrow G$,
  $h \leftarrow 0$\;
  \Repeat{no more real contractions}{
    $\C^h \leftarrow \mbox{Singletons}(G^h)$\;
    $\C^h \leftarrow$ \textsc{LM}$(G^h,C^h, \measure{intra}, \measure{inter}, \alpha)$\;
    $G^{h+1} \leftarrow \mbox{contract}(G^h, \C^h)$\;
    $h \leftarrow h+1$\;
  }
  \While{$h \geq 0$}{
      $h \leftarrow h-1$\;
      $\C^h \leftarrow \mbox{project}(\C^{h+1}, G^h)$\;
      $\C^h \leftarrow$ \textsc{LM}$(G^h, \C^h, \measure{inter}, \measure{intra}, \alpha)$\;
  }
 \Return{$\C^0$}\;
\end{algorithm}
\end{minipage}%
\hfil%
\begin{minipage}[b]{0.5\textwidth}
\begin{algorithm}[H]\footnotesize
 \caption{\textsc{Local Moving (LM)}}
 \label{algo:MOVE}
 \SetKwInOut{Input}{Input}\SetKwInOut{Output}{Output}
 \Input{graph $G$, clustering $\C_{\text{init}}$ of $G$, \measure{inter}, \measure{intra}, $\alpha$}
 \Output{clustering $\C$ of $G$}
\DontPrintSemicolon
 $\C \leftarrow$ $\C_{\text{init}}$\;
 \Repeat{no more changes}{
   \ForAll{$v \in V$}{
      $\mathcal{A} \leftarrow \{C \in \C \mid \measure{intra}(\moveVC{v}{C}) \geq \alpha$\;
       \hspace{1.5cm}$ \mbox{ and } |E(v, C)| > 0 \}$\;
      $N \leftarrow \arg\min\limits_{C \in \mathcal{A} \cup \{\}}\{\measure{inter}(\moveVC{v}{C})\}$ \;
      \If{$\measure{inter}(\moveVC{v}{N}) < \measure{inter}(\C)$}{
        move$(v, N)$\;
      }
    }
 }
 \Return{$\C$}\;
\end{algorithm}
\end{minipage}%
\end{myalgo}
\normalsize
\andreapar\textbf{Greedy Vertex Moving (GVM).} 
The key ingredient of GVM (Algo.~\ref{algo:GLM}) is a subprocedure that tries to greedily improve the objective function by letting vertices move to neighboring clusters (Algo.~\ref{algo:MOVE}).
This subprocedure repeatedly iterates through the vertex set and, for each vertex, performs the most improving move (subject to the constraint), potentially isolating a vertex, or leaving it where it was, until a local optimum is reached.
Starting with singletons, GVM first calls this subprocedure and contracts the resulting preliminary clustering into a super-graph, i.e.~each cluster becomes a vertex weighted with the number of vertices it represents, and edges are summarized such that edge weights reflect the number of edges in the original graph.
This whole process is iterated until local moving does not yield any further improvement, and results in a hierarchy of graphs with increasing coarseness.
In the second phase (refinement), the hierarchy is unfurled step by step by projecting the clustering of the $i+1$-th level of the hierarchy to level $i$, i.e.~the clusters in level $i$ are merged according to the clustering in level $i+1$.
After each step, LM is called again on the current level of the hierarchy to potentially improve the objective function further, until a clustering for the finest level, i.e.~the original graph, is obtained.  
\par
GVM is closely related to algorithms in the context of graph partitioning and has previously been used for modularity-based clustering without constraints \cite{bgll-f-08,rn-m-11}. 
Neither approximation guarantees nor subexponential bounds on the running time are known, but experimentally it has been shown to outperform the corresponding greedy merge algorithm with respect to both quality and efficiency.
For modularity, it can easily be shown that moving a vertex to a cluster it is not linked with is never the best choice, therefore it suffices to consider neighboring clusters.
Together with the observation that the change in modularity can be determined in constant time for each move if some information about the clustering is maintained, this yields a running time in $O(m)$ for each round in LM.
This latter observation on running time also holds for all intracluster density and intercluster sparsity measures except for \measure{mixd}, \measure{mixc} and \measure{mixe}, whose values are expensive to maintain.
\andreapar\textbf{Ensuring Strict Improvements.}
Another issue with a direct application of GVM to maximum-based measures is that iteratively traversing the whole vertex set is inefficient if only very few vertex moves potentially decrease the cut of the cluster with the currently worst value.
Even worse, if this cluster is not unique, it is likely that the search is stuck in a local minimum, as vertex moves generally can only improve the value for one of these cluster, not for all of them simultaneously.
If we try to prevent this by allowing vertex moves that are not strictly improving, we somehow have to ensure that the algorithm terminates after a finite number of operations.
We do this in a similar way as proposed in~\cite{gsw-dcgc-11b} for GM by greedily optimizing the lexicographical order of the intercluster sparsity values of all the clusters.
Let $L(\C) := \bigl(f(C_1), \ldots, f(C_k)\bigr), C_i \in \C,$ be the sequence of these values with decreasing intercluster density, i.e.~$(f(C_i) \geq f(C_{i+1})$ for $i \in \{1, \ldots, k-1\}$.
Then a clustering $\C$ is \emph{$L$-better} than $\C'$ if $L(\C)$ is lexicographically less than $L(\C')$.
We now determine for each vertex the set of clusterings that can be reached by moving it. 
If one of these clusterings is $L$-better than the current clustering,
the move that results in the $L$-best sequence is performed.
As we strictly improve the lexicographical order in each step, termination is guaranteed.
This means, we greedily optimize the maximum value but are also allowed to improve the intercluster sparsity of clusters more locally, yielding better efficiency and the possibility to escape local minima.
\andreapar\textbf{Determining the Best Move in $O(\deg(v))$ Time.}
It holds that any two clusterings resulting from leaving vertex $v$ untouched or from moving $v$ to a different (or new) cluster can be $L$-compared in constant time (see App. \ref{app:add_ex}).
Furthermore, it is immediate that moving a vertex to a cluster it is not linked to can never decrease the number of intercluster edges (\measure{nxe}).
This does not hold for \measure{gxd}, however, it is not hard to see that GVM never has to consider non-neighboring clusters for \measure{gxd} (see App.~\ref{app:add_ex}).
\begin{figure}
 	\centerline{\includegraphics[scale=0.8]{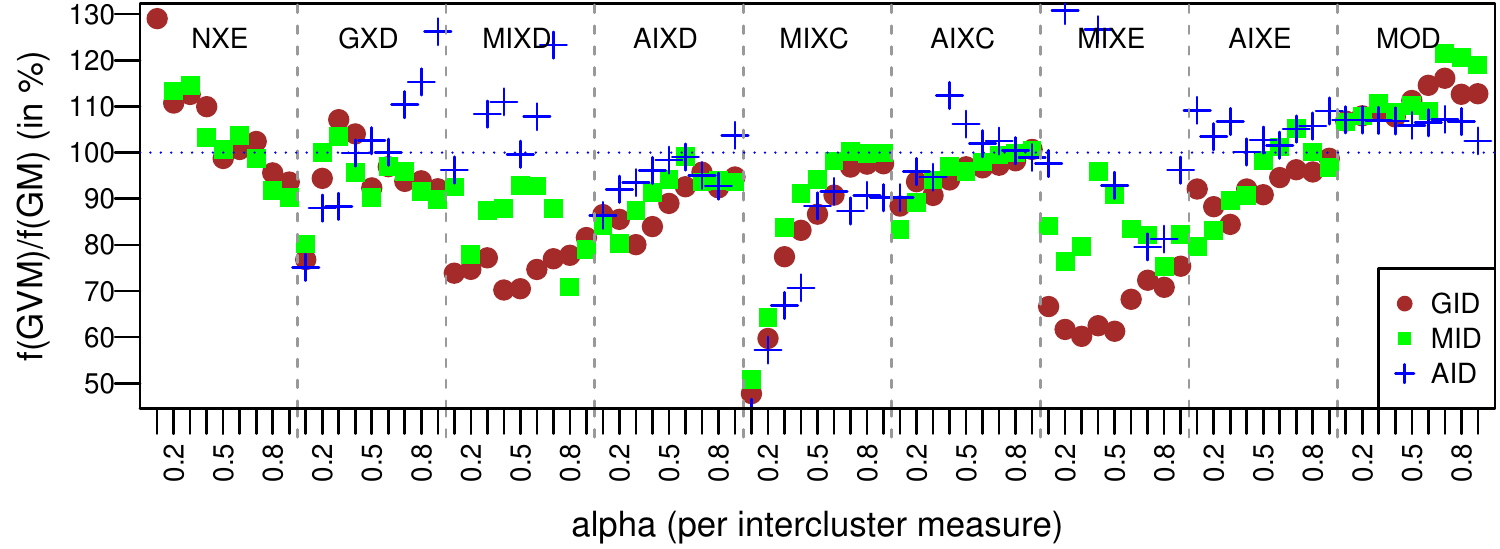}}
	\vspace{-1.5ex}
 	\caption{Qualitative comparison of GVM and GM.}
 	\label{fig:LM_vs_MOVE}
\end{figure}
For all other intercluster density measures this does not hold as can be seen in the examples in Fig.~\ref{fig:not_ddcmove:app} in App.~\ref{app:add_ex}. 
As configurations like these are only expected in degenerate cases, the impact on efficiency is large on sparse graphs, and unconnected clusters are not desirable in the context of graph clustering, we chose to restrict the set of feasible moves to neighboring clusters.
Together with the possibility to compare different moves in constant time, we get a time complexity of $O(m)$ for each round of the local move procedure for each of the combinations considered.
%
\section{Experiments}
\andreapar\textbf{Qualitative Comparison of Greedy Merge and Greedy Vertex Moving.}
Our first experiments address the question which flavor of greedy algorithm is better suited for \textsc{DCC}.
As test instances, we used all graphs listed in Table \ref{tab:real_world_graphs} with less than $1000$ vertices, these are real-world networks taken from the websites of Mark Newman~\cite{newmanWS}
and Alex Arenas~\cite{arenasWS} and are part of the clustering testbed of the 10th DIMACS Implementation Challenge \cite{dimacsWS}.
For all proposed combinations of measures,
Figure~\ref{fig:LM_vs_MOVE} shows the ratio of the intercluster density obtained by using GVM and GM, averaged over all graphs.
For modularity, this ratio is always greater than one, confirming that local moving yields better results, regardless of the choice and strength of the constraint.
In combination with \measure{gid} and \measure{mid}, this similarly holds for all other objectives except for \measure{nxe}, note that, in contrast to modularity, we aim to minimize these measures and therefore a value below one means that GVM attains better results.
For \measure{nxe}, the outcome depends on the value of $\alpha$ chosen.
In combination with \measure{aid}, the outcome is less clear, the results for \measure{nxe} are out of bounds as the ratio for some configurations exceeds 300 percents.
This can be explained by the observation that \measure{aid} happily allows (and thereby encourages) unbalanced clusterings, as bad intracluster density values of large clusters can easily be compensated by a set of small and dense clusters, and GM is known to have a tendency to produce unbalanced partitions.
As this most often leads to unintuitive clusterings, we deem \measure{aid} less suitable in the context of graph clustering.
Disregarding \measure{aid} for these reasons, in a vast majority of configurations, GVM outperforms GM. 
For tackling \textsc{DCC}, we thus solely use GVM, putting aside the algorithm based on greedy merging.
\par
\begin{figure}[t]
 	\centerline{\includegraphics[scale=0.9]{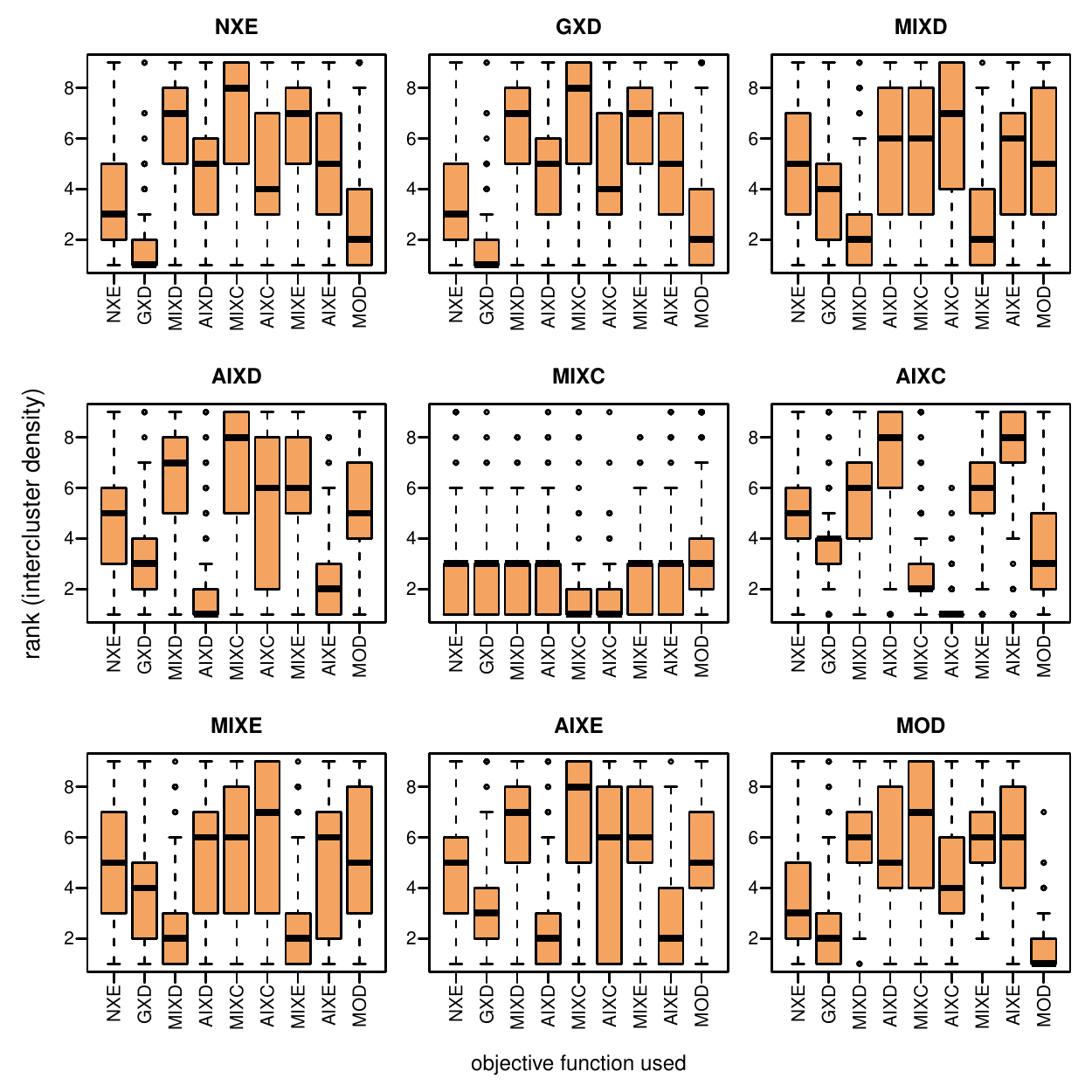}}
	\vspace*{-1.5ex}
 	\caption{Ranks for different intercluster density measures as objectives in the GVM-algorithm using \measure{gid} as constraint, evaluated by the intercluster density of the resulting clustering.}
 	\label{fig:gid}
\end{figure}

\begin{footnotesize}
\begin{table}
\footnotesize
    \begin{center}
	\begin{tabular}{|l|r|r||l|r|r|}
	\hline
	graph & n & m & graph & n & m\\
	\hline
	\hline
	karate(N) & 34 & 78 & netscience(N) & 1589 & 2742 \\
    dolphins(N) & 62 & 159 & power(N) & 4941 & 6594 \\
    lesmis(N) & 77 & 254 & hep-th(N) & 8361 & 15751 \\
    polbooks(N) & 105 & 441 & PGPgiantcompo(A) & 10680 & 24316\\
    adjnoun(N) & 112 & 425 & astro-ph(N) & 16706 & 121251\\
    football(N) & 115 & 616 & cond-mat(N) & 16726 & 47594\\
    jazz(A) & 198 & 2742 & as-22july06(N) & 22963 & 48436\\
    celegansneural(N) & 297 & 2148 & cond-mat-2003(N) & 31163 & 120029\\
    celegans\_metabolic(A) & 453 & 2039 & cond-mat-2005(N) & 40421 & 175693\\
    polblogs(N) & 1490 & 16718 &  &  &\\
	\hline
	\end{tabular}
	\end{center}
	\caption{List of the real world test instances ordered by increasing number of vertices. These are taken from the webpages of Arenas(A) \cite{arenasWS} and Newman(N) \cite{newmanWS} and are often used to compare clustering algorithms. All graphs are part of the clustering testbed of the 10th DIMACS Implementation Challenge \cite{dimacsWS}.}
	\label{tab:real_world_graphs}
\end{table}
\end{footnotesize}
\normalsize
\andreapar\textbf{Effectiveness of Different Objective Functions.}
The next question we pose is, if each of the intercluster density measures is effective in optimizing itself when used as \measure{inter} in GVM. 
To answer this question, we conducted the following experiment on the set of
graphs listed in Table \ref{tab:real_world_graphs}.
In the following, let $\text{GVM}_{i, \alpha, x}$ denote GVM incorporating the constraint $i(\C) \geq \alpha$ and the objective $x(\C)$.
For each setup of DCC, i.e.~intracluster measure $i$, intercluster measure $x$ and $\alpha \in \{0.0, 0.1, \ldots, 1.0\}$, we ranked the clusterings obtained by $\text{GVM}_{i, \alpha, y}$ by their performance with respect to $x$, using all possible objectives $y$ for GVM.
Figure \ref{fig:gid} shows the distribution of these ranks over all configurations involving \measure{gid}, grouped by $x$. 
The outcome of this experiment is less clear than what might be expected---none of the intercluster measures, not even modularity, scores the best quality with respect to itself in all configurations.
Nonetheless, in general, except for \measure{nxe} which is clearly dominated by \measure{gxd}, each objective optimizes itself quite well.
This also holds for \measure{mid}, while for \measure{aid}, the outcome is even less clear, as can be seen in Figures~\ref{fig:mid}, \ref{fig:aid} in App.~\ref{app:add_effectiveness}.
\andreapar\textbf{Reference Algorithms.}
For a more comprehensive assessment of GVM as a means to address \textsc{DCC}, we use the following reference algorithms:
\vspace*{-0.5ex}
\begin{itemize}
    \item \emph{Iterative Conductance Cutting (ICC)}~\cite{kvv-cgds-00}:
    This top-down algorithm iteratively splits the input graph into two subgraphs based on a cut with low conductance.
    The process stops when the conductance of the cut exceeds a given threshold, which we set to $0.4$ in our experiments.
	\item \emph{Markov-Clustering (MCL)}~\cite{phd-dongen-02}:
	Emulating a random walk, the matrix of transition probabilities is alternately taken to to the power of $e$ and renormalized after taking each entry to the power of $r$, where $e$ and $r$ are input parameters.
In our experiments, we set $r$ and $e$ to $2$.
	\item \emph{Geometric MST Clustering (GMC)}~\cite{bgw-egca-03}: First, a spectral embedding of the graph in $d$-dimensional space is built.
	Then the algorithm constructs a Euclidean minimum spanning tree and successively deletes the heaviest edge.
	This defines a sequence of forests whose connected components induce a set of clusterings.
	Among these clusterings, the one with the best value according to some given objective function is chosen.
	\item \emph{Multi-Level Modularity (MOD)}~\cite{rn-m-11}: This is the GVM-algorithm based solely on modularity without using any constraint. This algorithm has been shown to perform very well in the context of Modularity optimization~\cite{rn-m-11}.
\end{itemize}
\vspace*{-0.5ex}
\andreapar\textbf{Comparison Based on Intracluster Density Found by Reference Algorithms.}
ICC, MCL and MOD do not incorporate constraints on the intracluster density of the resulting clustering.
Nonetheless, it is still possible to evaluate them with respect to those variants of DCC, where $\alpha$ is set to the intracluster density found by these algorithms.
In other words, given the same constraint a reference algorithm $\mathcal{A}$ implicitly adheres to, how well does GVM compare to $\mathcal{A}$ wrt.~DCC?
\par
We first ran ICC, MCL and MOD on all test instances in Table~\ref{tab:real_world_graphs} and recorded the intracluster density values of the resulting clusterings. 
Then, for each reference algorithm $\mathcal{A}$, $i$, recorded corresponding intracluster density $\alpha$ and $x$, we compare the clustering obtained by $\text{GVM}_{i, \alpha, x}$ to the clustering of $\mathcal{A}$ with respect to $x$. 
For GMC the experiments slightly differ as GMC requires an objective function.
We filled this degree of freedom by choosing $f(\C)=i(\C)-x(\C)$ as the objective function for the experiments using $i$ as intracluster and $x$ as intercluster density measure.
This seemed to be the fairest way of comparison and in almost all cases led to non-trivial clusterings.
\par
\begin{table}[b]
\footnotesize{
\begin{center}
\begin{tabular}{|l|rrrr|rrrr|rrrr|}
\hline
 & \multicolumn{4}{c|}{\measure{gid}} & \multicolumn{4}{c|}{\measure{mid}} & \multicolumn{4}{c|}{\measure{aid}} \\
  & \multicolumn{1}{c}{\measure{ICC}} & \measure{MCL} & \measure{MOD} & \measure{GMC} & \measure{ICC} & \measure{MCL} & \measure{MOD} & \measure{GMC} & \measure{ICC} & \measure{MCL} & \measure{MOD} & \measure{GMC}\\
  \hline
\measure{nxe} & 84 &  95 &  16 &  63 &  89 &  95 &  63 &  74 &  95 & 100 & 100 &  63 \\
  \measure{gxd} & 84 & 100 &  42 & 100 &  95 & 100 &  84 & 100 &  95 & 100 & 100 &  84 \\
  \measure{aixd} & 84 & 100 &  42 & 100 &  89 & 100 &  37 &  95 &  95 & 100 & 100 &  84 \\ 
  \measure{aixc} & 84 & 100 &  21 &  53 &  95 & 100 &  79 &  42 &  95 &  95 & 100 &  63 \\ 
  \measure{aixe} & 84 &  95 &  42 &  89 &  89 &  95 &  42 &  95 &  95 &  95 &  95 &  95 \\ 
  \measure{mixd} & 84 &  95 &  53 &  84 &  89 & 100 &  74 &  89 &  89 &  95 &  89 &  74 \\ 
  \measure{mixc} & 89 &  95 &  42 &  37 &  89 &  95 &  63 &  37 &  89 &  95 &  84 &  21 \\ 
  \measure{mixe} & 89 &  95 &  58 &  89 &  84 &  95 &  47 &  79 &  95 &  95 &  89 &  63 \\ 
   \hline
\end{tabular}
\end{center}
}
\caption{Comparison of GVM and reference algorithms. 
 Entries represents the percentage of graphs GVM compares favorably.}
\label{tab:nat_alphas_percentage}
\end{table}

\begin{table}
\footnotesize{
\begin{center}
\begin{tabular}{|l|rrrr|rrrr|rrrr|}
\hline
 & \multicolumn{4}{c|}{\measure{gid}} & \multicolumn{4}{c|}{\measure{mid}} & \multicolumn{4}{c|}{\measure{aid}} \\
  & \multicolumn{1}{c}{\measure{ICC}} & \measure{MCL} & \measure{MOD} & \measure{GMC} & \measure{ICC} & \measure{MCL} & \measure{MOD} & \measure{GMC} & \measure{ICC} & \measure{MCL} & \measure{MOD} & \measure{GMC}\\
  \hline
\measure{nxe} & 0.67 & 0.52 & 1.17 & 1.26 & 0.42 & 0.08 & 0.97 & 0.88 & 0.03 & 0.06 & 0.05 & 8.07 \\ 
  \measure{gxd} & 0.64 & 0.50 & 1.07 & 0.11 & 0.40 & 0.09 & 0.89 & 0.10 & 0.07 & 0.10 & 0.13 & 0.76 \\ 
  \measure{aixd} & 0.47 & 0.32 & 5.30 & 0.25 & 0.34 & 0.06 & 5.08 & 0.23 & 0.18 & 0.12 & 0.22 & 0.61 \\ 
  \measure{aixc} & 0.57 & 0.29 & 2.17 & 0.28 & 0.39 & 0.05 & 0.81 & 0.27 & 0.41 & 0.27 & 0.37 & 7.87 \\
  \measure{aixe} & 0.49 & 0.39 & 5.55 & 0.31 & 0.36 & 0.14 & 5.22 & 0.31 & 0.19 & 0.13 & 0.24 & 1.45 \\ 
  \measure{mixd} & 0.45 & 0.34 & 0.96 & 0.41 & 0.39 & 0.07 & 1.27 & 0.30 & 0.21 & 0.18 & 0.32 & 3.17 \\
  \measure{mixc} & 0.69 & 0.58 & 1.15 & 0.34 & 0.47 & 0.15 & 1.09 & 0.30 & 0.44 & 0.39 & 0.46 & 1.60 \\ 
  \measure{mixe} & 0.48 & 0.26 & 1.25 & 0.57 & 0.39 & 0.14 & 1.28 & 0.63 & 0.13 & 0.16 & 0.28 & 3.02 \\ 
   \hline
\end{tabular}
\end{center}
}
\caption{Comparison of GVM and reference algorithms. 
 Entries represent the mean ratio of the respective intercluster measure $x$ obtained by $\text{GVM}$ and reference algorithm.}
\label{tab:nat_alphas_ratio}
\end{table}
Table \ref{tab:nat_alphas_percentage} and Table \ref{tab:nat_alphas_ratio} show the percentage of graphs where the greedy algorithm for $x$ compares favorably and the arithmetic mean of the ratio of $x$ obtained with GVM and with the reference algorithm.
As we aim to minimize intercluster density, a value below $1$ indicates that the greedy algorithm succeeds in beating the reference algorithm and vice versa.
Compared to ICC and MCL, GVM clearly yields better results.
The same holds for GMC, except if used in combination with \measure{aid}, where GMC sometimes produces far better results.
This can be explained by the fact that \measure{aid} does not punish unbalancedness and GMC naturally leads to very unbalanced clusterings in most instances.
The outcome of the comparison with the modularity-based algorithm is less clear.
For \measure{aid}, GVM performs better, which is not surprising as modularity strongly discourages unbalanced clusterings.
For \measure{mid}, GVM still beats MOD in the majority of configurations, while for \measure{gid}, this only holds for slightly less than half of the configurations.
Furthermore, it is worth mentioning that especially for \measure{aixd} and \measure{aixe} there are instances where modularity minimizes these functions far better than the respective greedy algorithms.
Altogether, the comparison with ICC, MCL and GMC suggests that GVM effectively addresses DCC, while the comparison with MOD shows that optimizing modularity is similarly effective in minimizing cut-based intercluster sparsity measures.
\setlength{\tabcolsep}{2pt}
\andreapar\textbf{Recovering Planted Partitions.} 
To compare the different objective functions qualitatively, we evaluated how well the corresponding GVM-algorithms are able to reconstruct planted partitions in random graphs. 
As a comparison, we also give the results obtained by MOD.
Due to higher running times and large numbers of experiments, we omit a comparison with ICC, MCL and GMC.

\andreapar\textbf{Random Graphs Generated.}
We use an adapted Erd\H{o}s-R\'{e}nyi-model, where, starting from a given reference partition, the probability that vertices in the same set (in different sets) are connected equals $p_{in}$ ($p_{out}$). The number of vertices ($n$) and clusters ($k$) as well as the skewness of the distribution of cluster sizes ($\beta$) of the planted partition are input parameters.
Setting $\beta$ to $1.0$ corresponds to uniform cluster sizes, values below and above $1$ cause this distribution to be skewed, for more details see \cite{gs-agdcr-09}. 
As configurations, we fixed $n$ to $10000$ and chose $p_{in}$ and $p_{out}$ such that the average number of intracluster (intercluster) edges a vertex is incident to equals $5$ ($3$). 
To determine the reference partition, we used all combinations of $k \in \{10, 100, 300\}$ and $\beta \in \{0.3, 1.0, 2.0\}$.
For each configuration, we generated $100$ instances and always averaged obtained values. 
\andreapar\textbf{Distance Measures.}
To compare the clusterings obtained with the different algorithms with the reference clustering, we use the following graph-based distance measures taken from~\cite{dggw-ecgc-08}:
\vspace{-1ex}
\begin{itemize}
  \item \emph{Graph-based Rand Index ($R_g$)}: Let $\C_1$ and $\C_2$ be clusterings and $e_{11}$ ($e_{00}$) the number of edges which are intracluster (intercluster) wrt.\ both $\C_1$ and $\C_2$.
 Then, $R_g(\C_1, \C_2) = 1-(e_{11} + e_{00})/m $.
  \item \emph{Editing Set Difference ($ESD$)}: For a clustering $\C$, its editing set $F_{\C}$ is the set of edges requiring insertion or removal such that the clusters in $\C$ form disjoint cliques. 
Then, for clusterings $\C_1$ and $\C_2$, their editing set difference is defined as
		$ESD(\C_1, \C_2) = 1-|F_{\C_1} \cap F_{\C_2}|/|F_{\C_1} \cup F_{\C_2}|.$
\end{itemize}
\vspace{-1ex}
\andreapar\textbf{Parameters and Evaluation.}
As an exhaustive parameter search 
for all configurations would be far too expensive, we always set $\alpha$ to $75$ percent of the expected global intracluster density $p_{\text{in}}$.
We deemed taking the actual value of $p_{\text{in}}$ too strict, as, especially for \measure{mid}, even the reference clustering of the generator most likely does not meet this constraint.    
The previous experiments indicate that there are configurations where particular objective functions used in GVM do not score the best results with respect to themselves. 
As our goal is to compare good clusterings with respect to different combinations of $i$ and $x$, independent of artifacts of GVM, we chose the following approach:
For a combination $i$, $\alpha$, $x$, we evaluated the clustering that, among all results obtained with GVM using $i \geq \alpha$ as constraint, is best with respect to $x$ (as opposed to simply evaluating $\text{GVM}_{i, \alpha, x}$).
Furthermore, preliminary experiments confirmed that constraining \measure{aid} leads to very unintuitive and unbalanced clusterings, which is mirrored by the fact that the corresponding versions of \textsc{DCC} are far less effective in finding the hidden clustering.
We hence excluded \measure{aid} in the discussion of the results.
\begin{figure}[htbp]
  \subfloat{%
    \begin{minipage}[t]{\textwidth}%
      \centerline{\includegraphics[scale=0.8]{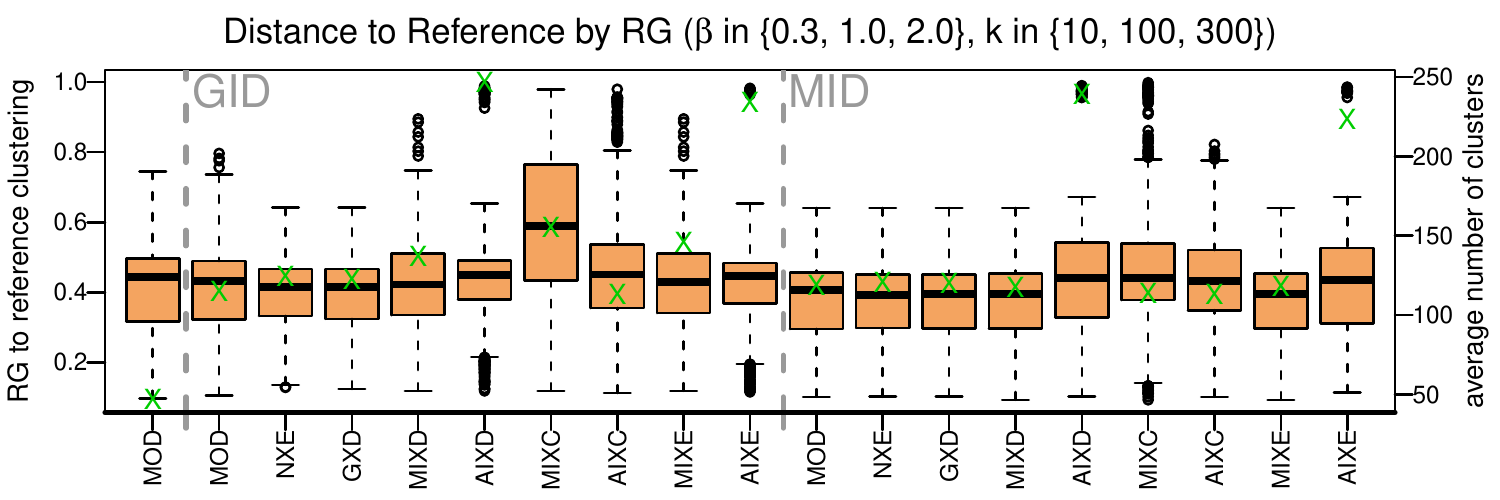}}%
    \end{minipage}%
  }%
  \hfill%
  \vspace{0.1cm}
  \subfloat{%
    \begin{minipage}[t]{\textwidth}%
      \centerline{\includegraphics[scale=0.8]{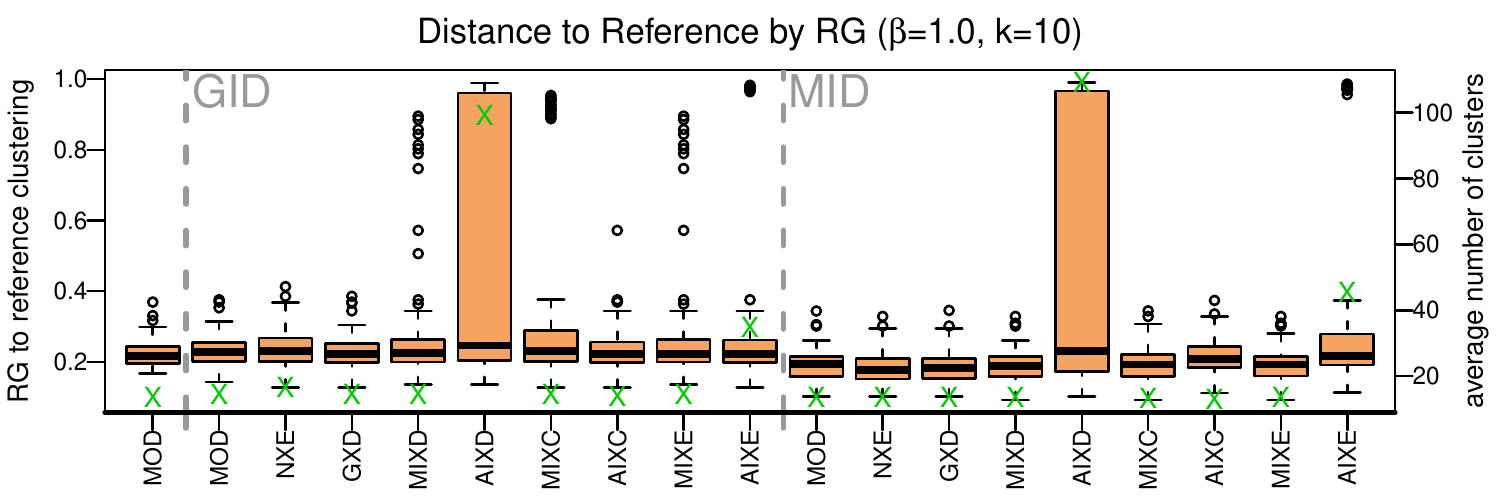}}%
    \end{minipage}%
  }%
  \hfill%
  \vspace{0.1cm}
  \subfloat{%
    \begin{minipage}[t]{\textwidth}%
      \centerline{\includegraphics[scale=0.8]{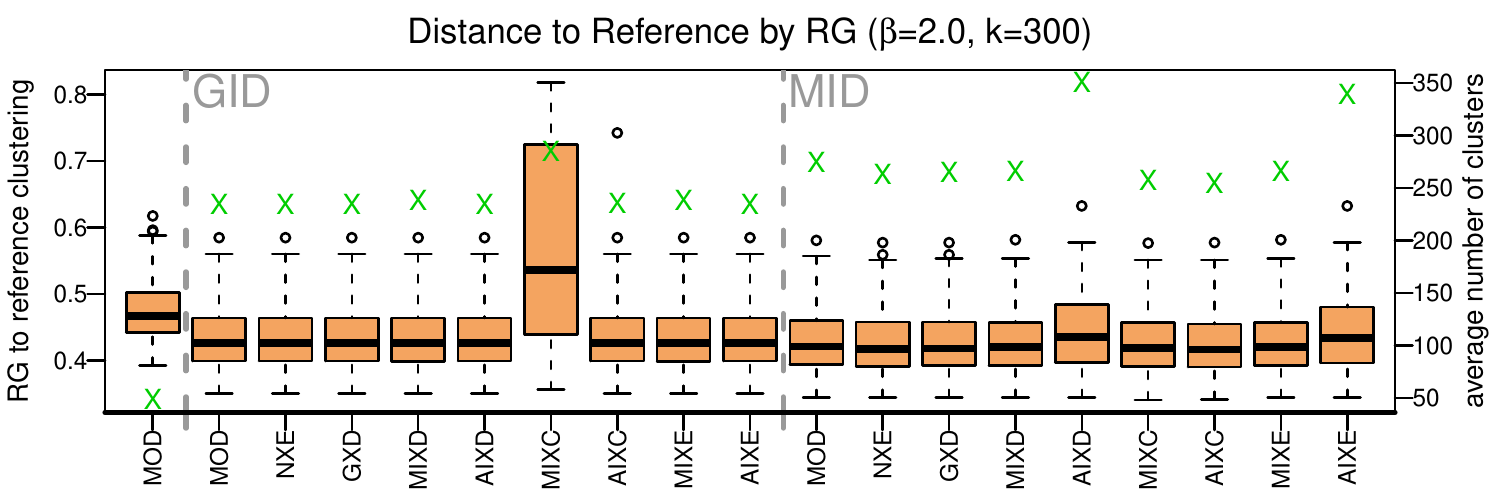}}%
    \end{minipage}%
  }%
    \hfill%
  \vspace{0.1cm}
  \subfloat{%
    \begin{minipage}[t]{\textwidth}%
      \centerline{\includegraphics[scale=0.8]{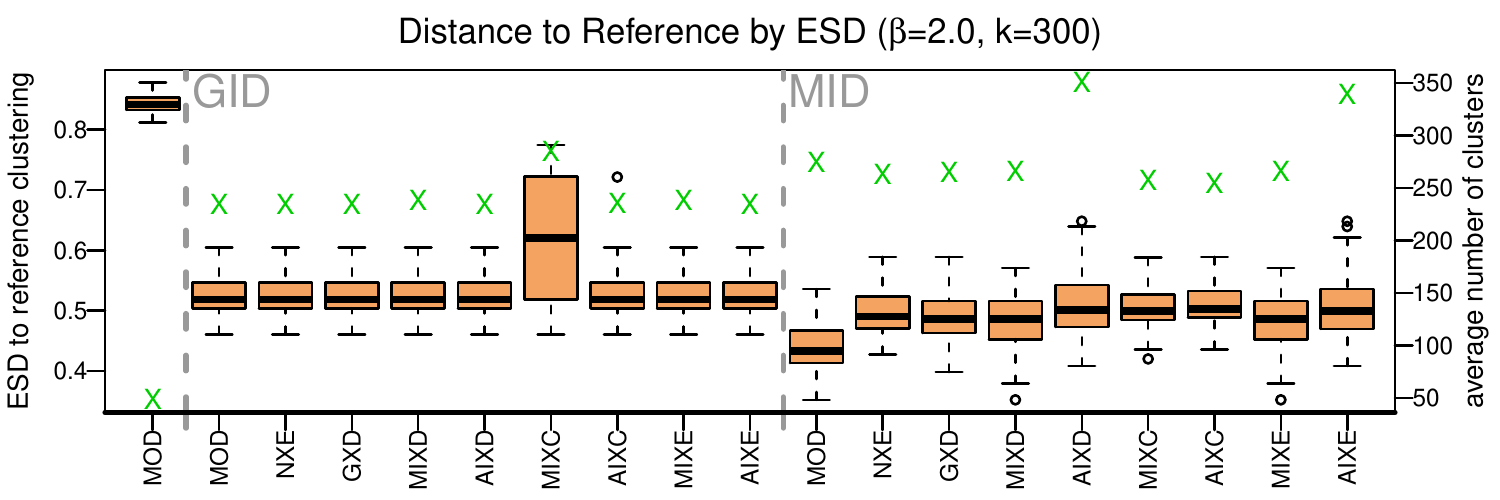}}%
    \end{minipage}%
  }%
  \vspace*{-1ex}
  \caption{Distance to reference clustering (boxplots, left-hand $y$-axis) and number of clusters discovered in planted partition graphs (green $\times$-marks, right-hand $y$-axis), different configurations}%
  \label{fig:Gnp}%
\end{figure}
%
\andreapar\textbf{Results on Planted Partition Graphs.}
Figure~\ref{fig:Gnp} shows the results for selected configurations, the results for the whole set of experiments can be found in App.~\ref{sec:Gnp_add}.
In the first plot 
it can be seen that, in general, the clusterings that are ranked best with respect to \measure{mod}, \measure{nxe} and \measure{gxd} are most similar to the reference. 
%
\par\emph{Constraining modularity by \measure{mid} improves its results.}
This especially holds for the experiments with high skewness ($\beta=2$) and $k=300$.
In these experiments, modularity finds far less clusters than expected, partially due to its known resolution limit~\cite{bf-rlcd-07}, which can be circumvented by steering the coarseness of the clustering by constraining the intracluster density.
Another interesting fact is that ESD punishes these coarse clusterings far more than $R_g$. 
%
\par\emph{Fine reference clusterings disbalance maximum objectives.}
Compared to the above,
especially \measure{mixc} in combination with \measure{gid} yields worse similarity values.
This, and the slightly increased cluster count can be explained by a tendency of \measure{mixc} to favor unbalanced clusterings if the expected number of clusters is high ($k=300$), which also explains why this effect does not happen in combination with \measure{mid} that does not allow very unbalanced clusterings.
To a smaller extent, the same observation also holds for the other maximum measures, 
as can be seen for $k=300$ and $\beta=1.0$.
%
\par\emph{\measure{aixe} and especially \measure{aixd} identify many clusters.}
Another striking observation is that the average number of clusters in clusterings found by \measure{aixd} and \measure{aixe}, indicated by the green $\times$-marks, is much higher than the average number of clusters in the reference.
This especially stems from the experiments with few clusters.
In the configuration with $\beta=1$ and $k=10$, it can also be seen that these measures differ the more, the coarser the expected clustering gets.
This is not unexpected, as the denominator of \measure{aixd} grows more slowly with the number of vertices in the cluster than the denominator of \measure{aixe}, meaning that \measure{aixd} is less eager to produce very large clusters.
Additionally, in \cite{gsw-dcgc-11b} it was proven that with the exception of \measure{aixd}, all intercluster measures considered here can always be ameliorated by merging two existing clusters (unboundedness), which is also a hint that \measure{aixd} is less likely to produce coarse clusterings than the other measures.
\andreapar\textbf{Implementation and Running Times.}
The algorithms ICC, MCL, GMC and GM are implemented in Java 1.6.0\_22 using the graph library yFiles \cite{yfiles}.
GVM (also incorporating MOD as a special case) is implemented in C++ using version 1\_42 of the Boost Graph Library \cite{boost} and compiled with gcc 4.5.2 with optimization level 4.
The focus of this evaluation is on the quality of the resulting clusterings, not on running times.
However, to get a rough impression about the latter, clustering cond-mat-2005 on a 2.1 GHz AMD Opteron processor takes about 6 hours with ICC, 1 hour and 50 minutes with MCL, 5 minutes with GMC and 3 to 15 seconds with GVM, depending on the parameter setting.
With our prototype implementation (not including the improvements proposed in \cite{gsw-dcgc-11b}) of GM, clustering the much smaller celegans\_metabolic takes over 2 minutes.
\section{Conclusion}
This work is an experimental evaluation of algorithms for the optimization problem \textsc{Density}-\textsc{Constrained Clustering} (DCC).
We first evaluated two greedy heuristics, vertex moving and cluster merging, against each other and against algorithms from the literature.
Vertex moving proved reliably superior to cluster merging and, in many cases, beats the results of the reference algorithms.
Our results also show that a well-known modularity-based algorithm implicitly addresses DCC quite well, revealing similarities between cut-based intercluster sparsity measures and modularity.
In the second part, we addressed the question whether different combinations of intracluster density and intercluster sparsity measures are suitable to guide algorithms in recovering planted partitions in random graphs. 
The results suggest that minimizing the average intercluster expansion or density of the clusters overestimates the number of clusters if the expected clustering is coarse, while the maximum intercluster measures lead to unbalanced clusters if the expected clustering is fine and the constraint on the intracluster density does not force the clustering to be balanced.
Additionally, it can be seen that the known resolution limit for modularity can be circumvented if the coarseness of the clustering is controlled by an additional constraint on the intracluster density of the clustering.

\bibliographystyle{splncs03}
\bibliography{references.bib,myRefs.bib}
\newpage
\appendix
\section{Additional Examples and Explanations}
\label{app:add_ex}

\textbf{Maximum Functions: Clusterings resulting from vertex moves can be $L$-compared in constant time.}
\emph{%
For three distinct clusters $C$, $A$ and $B$ in $\C$ and $v \in C$ it holds that:
\begin{itemize}
  \item $\moveVC{v}{A}$ is $L$-better than $\C \Leftrightarrow \bigl\{C \setminus \{v\}, A \cup \{v\} \bigr\}$ is $L$-better than $\{C, A\}$
  \item $\moveVC{v}{A}$ is $L$-better than $\moveVC{v}{B} \Leftrightarrow \bigl\{A \cup \{v\} , B \bigr\}$ is $L$-better than $\bigl\{B \cup \{v\}, A \bigr\}$
\end{itemize}
}
If the volume, size and number of out-going edges of the clusters $A$, $B$ and $C$ are maintained by the algorithm, the density/conductance/expansion of $C, A, B, C \setminus \{v\}, A \cup \{v\}$ and $B \cup \{v\}$ can be determined in constant time.
Hence, the conditions on the right-hand side can be evaluated in constant time, which can be used to determine the best move for a vertex efficiently. 
\smallskip
\andreapar\textbf{Connectedness of \measure{gxd}.}
The following equation shows that GVM never has to consider non-neighboring clusters for \measure{gxd}, as isolating the respective vertex is always more beneficial.
Let $v \in V$, $A := C(v) \setminus \{v\}$ and $B \in \C$ such that $m_{\{v\}, B} = 0$, then:
\begin{align*}\label{eq:gxd_connected}
\measure{gxd}(\moveVC{v}{\{\}}) &= \frac{\sum_{C_i, C_j, j>i} m_{C_i, C_j} + m_{\{v\}, A}}{\sum_{C_i, C_j, j>i}|C_i||C_j| + |A|}\\
&<\frac{\sum_{C_i, C_j, j>i} m_{C_i, C_j} + m_{\{v\}, A} - \overbrace{m_{\{v\}, B}}^{=0}}{\sum_{C_i, C_j, j>i}|C_i||C_j| + |A| - \underbrace{|B|}_{> 0}}\!=\!\measure{gxd}({\moveVC{v}{B}})\\
\end{align*}
\begin{figure}[htbp]
  \subfloat[\measure{mixd}, \measure{mixe}]{%
    \begin{minipage}[t]{.3\textwidth}%
      \centerline{\includegraphics[scale=1]{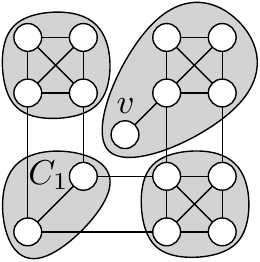}}%
      \label{fig:mpxc_not_ddcmove:small}%
    \end{minipage}%
  }%
  \hfill%
  \subfloat[\measure{mixc}]{%
    \begin{minipage}[t]{.3\textwidth}%
      \centerline{\includegraphics[scale=1]{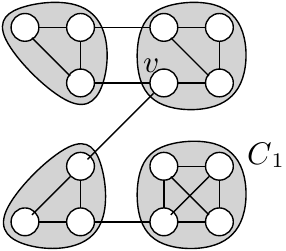}}%
      \label{fig:mixc_not_ddcmove:small}%
    \end{minipage}%
  }%
  \hfill
  \subfloat[\measure{aixc}, \measure{aixe}, \measure{aixd}]{%
    \begin{minipage}[t]{.4\textwidth}%
      \centerline{\includegraphics[scale=1]{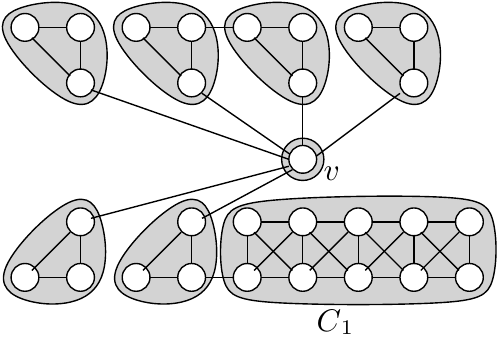}}%
      \label{fig:aixc_not_ddcmove:small}%
    \end{minipage}%
  }%
  \caption{Examples illustrating that most measures considered do not enforce connected moves. Given the clusterings indicated by the gray areas, among all moves involving $v$, moving $v$ to cluster $C_1$ yields the largest decrease in the objective function.}%
  \label{fig:not_ddcmove:app}%
\end{figure}

\clearpage

\section{Effectiveness of Different Objective Functions: Additional Plots}

\label{app:add_effectiveness}
\begin{figure}[b]
 	\centerline{\includegraphics[scale=0.9]{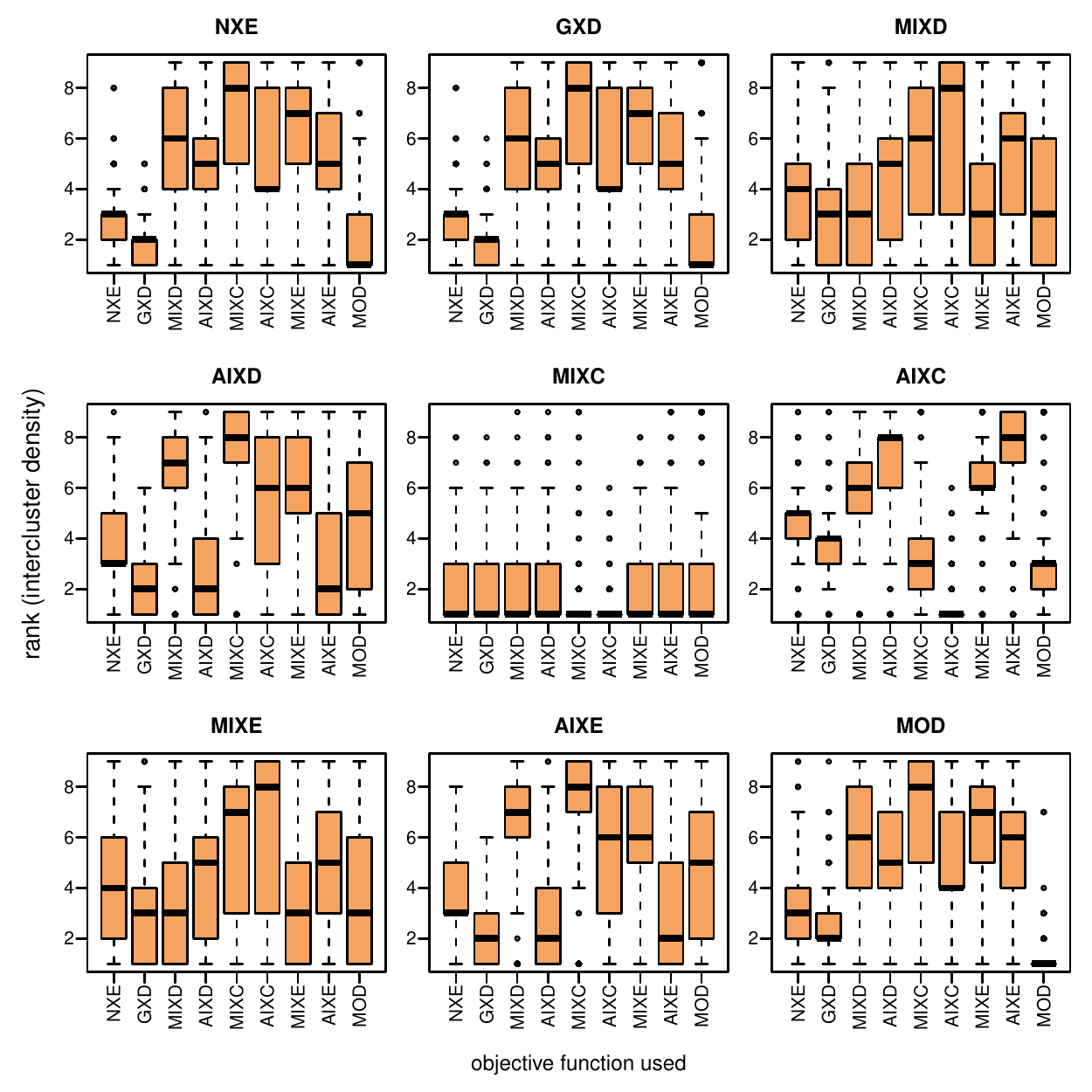}}
 	\caption{Ranks for different intercluster density measures as objectives in the GVM-algorithm using \measure{mid} as constraint, evaluated by the intercluster density of the resulting clustering.}
 	\label{fig:mid}
\end{figure}

\begin{figure}[b]
 	\centerline{\includegraphics[scale=0.9]{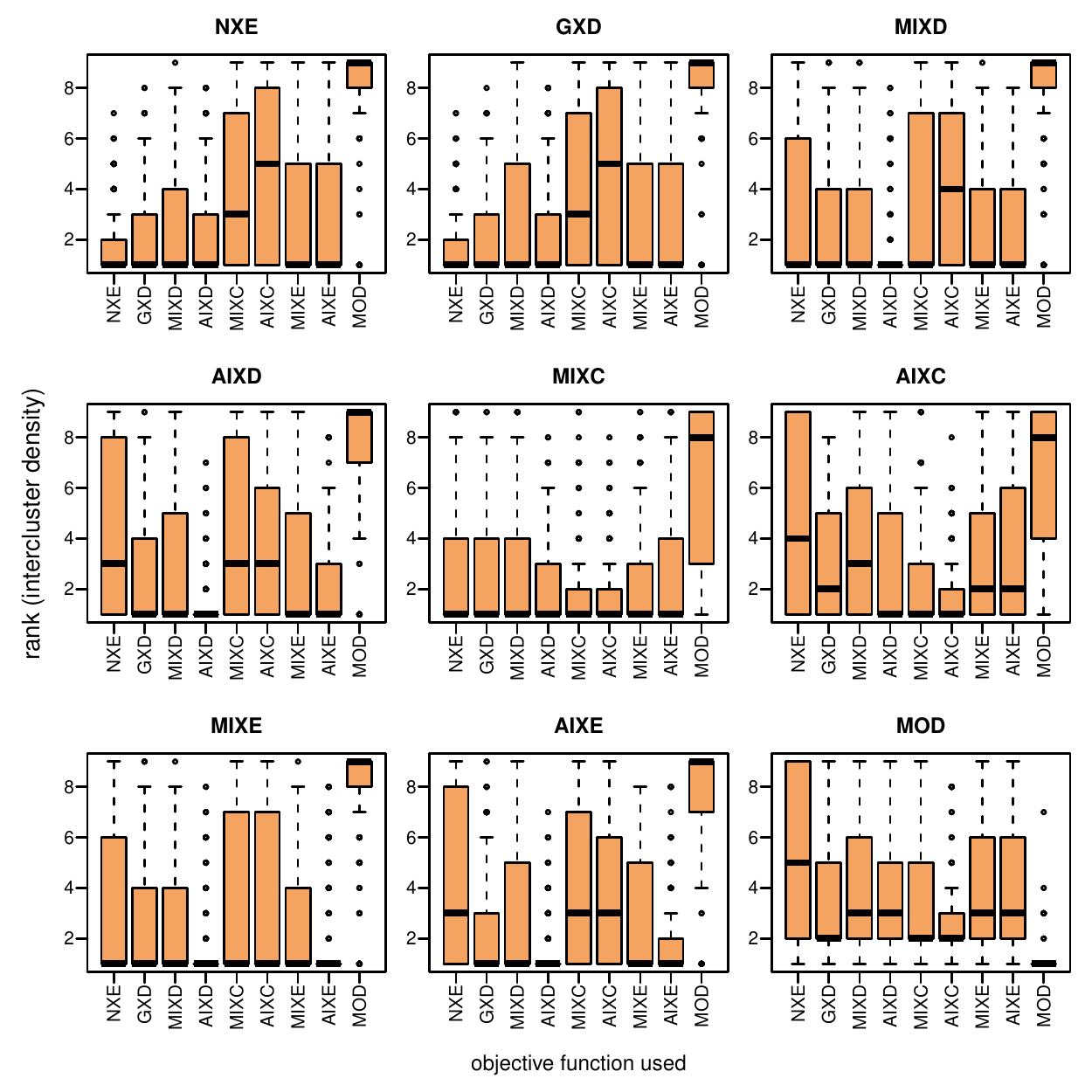}}
 	\caption{Ranks for different intercluster density measures as objectives in the GVM-algorithm using \measure{aid} as constraint, evaluated by the intercluster density of the resulting clustering.}
 	\label{fig:aid}
\end{figure}

\clearpage

\section{Complete Experiments with Planted Partition Graphs}
\label{sec:Gnp_add}
\begin{figure}[htbp]
  \subfloat{%
    \begin{minipage}[t]{\textwidth}%
      \centerline{\includegraphics[scale=0.8]{distancesToRef_G_ARAND_betaall_kall}}%
    \end{minipage}%
        }%
  \hfill%
  \vspace{0.0cm}
  \subfloat{%
    \begin{minipage}[t]{\textwidth}%
      \centerline{\includegraphics[scale=0.8]{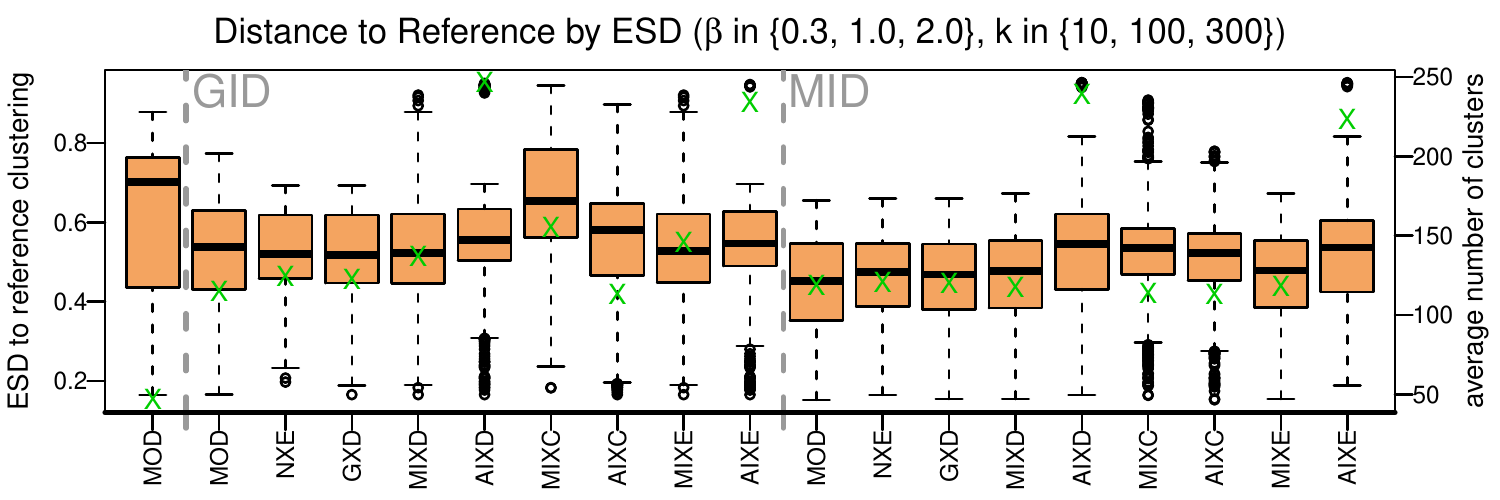}}%
    \end{minipage}%
  }%
  \hfill%
  \vspace{0.0cm}
  \subfloat{%
    \begin{minipage}[t]{\textwidth}%
      \centerline{\includegraphics[scale=0.8]{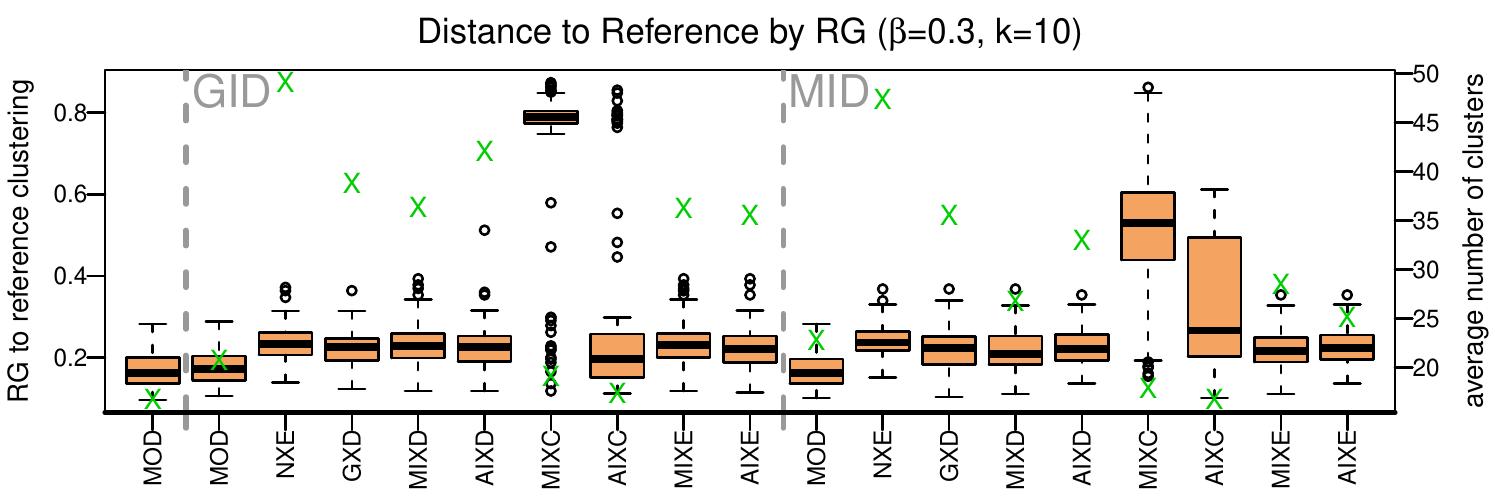}}%
    \end{minipage}%
  }%
    \hfill%
  \vspace{0.0cm}
  \subfloat{%
    \begin{minipage}[t]{\textwidth}%
      \centerline{\includegraphics[scale=0.8]{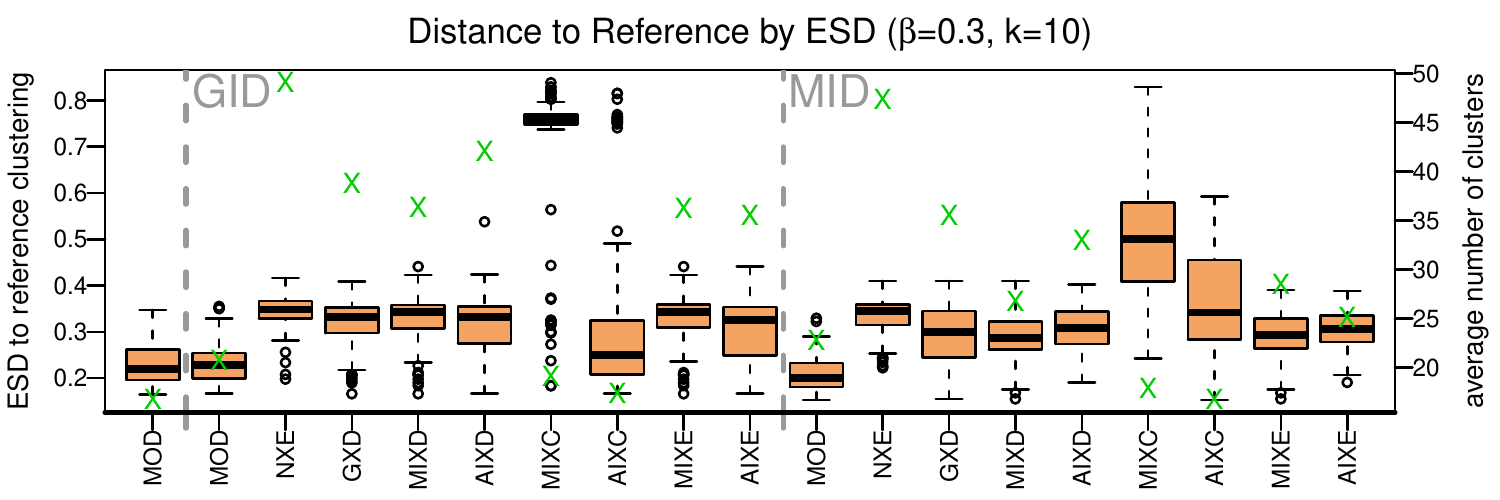}}%
    \end{minipage}%
  }%
\end{figure}

\begin{figure}[t]
  \subfloat{%
    \begin{minipage}[t]{\textwidth}%
      \centerline{\includegraphics[scale=0.8]{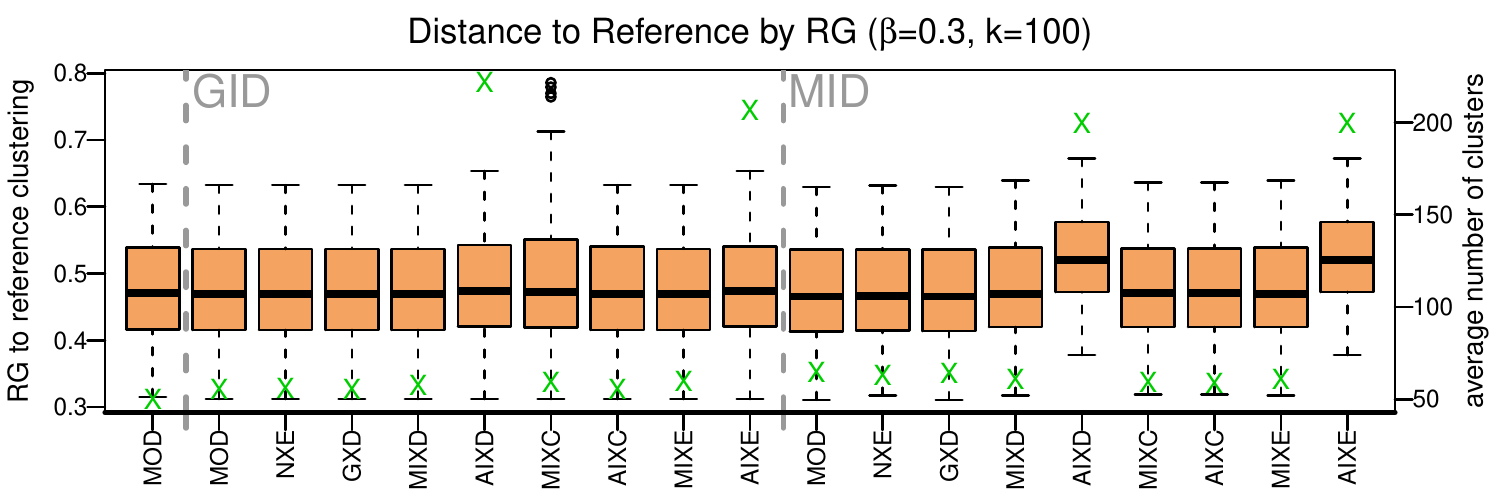}}%
    \end{minipage}%
        }%
  \hfill%
  \vspace{0.0cm}
  \subfloat{%
    \begin{minipage}[t]{\textwidth}%
      \centerline{\includegraphics[scale=0.8]{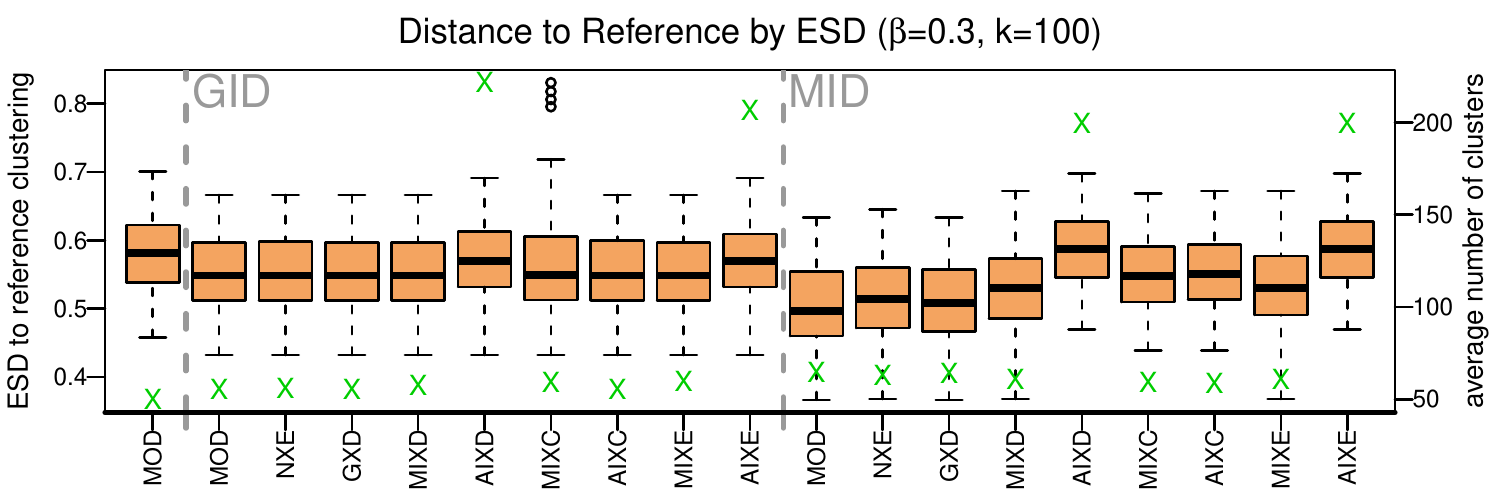}}%
    \end{minipage}%
  }%
  \hfill%
  \vspace{0.1cm}
  \subfloat{%
    \begin{minipage}[t]{\textwidth}%
      \centerline{\includegraphics[scale=0.8]{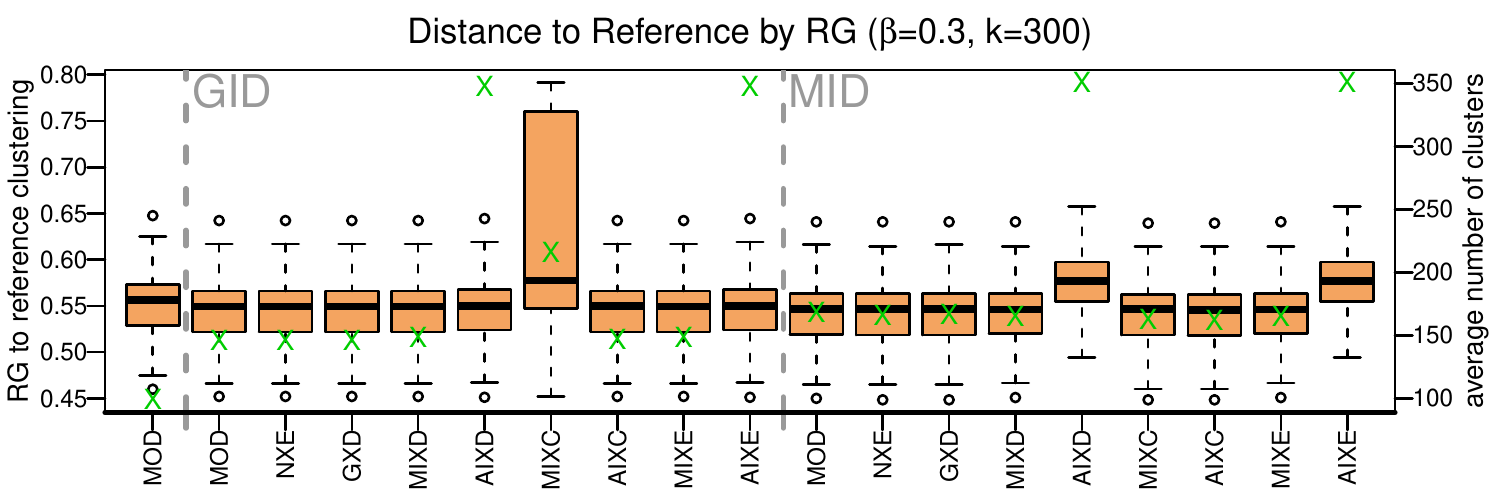}}%
    \end{minipage}%
  }%
    \hfill%
  \vspace{0.1cm}
  \subfloat{%
    \begin{minipage}[t]{\textwidth}%
      \centerline{\includegraphics[scale=0.8]{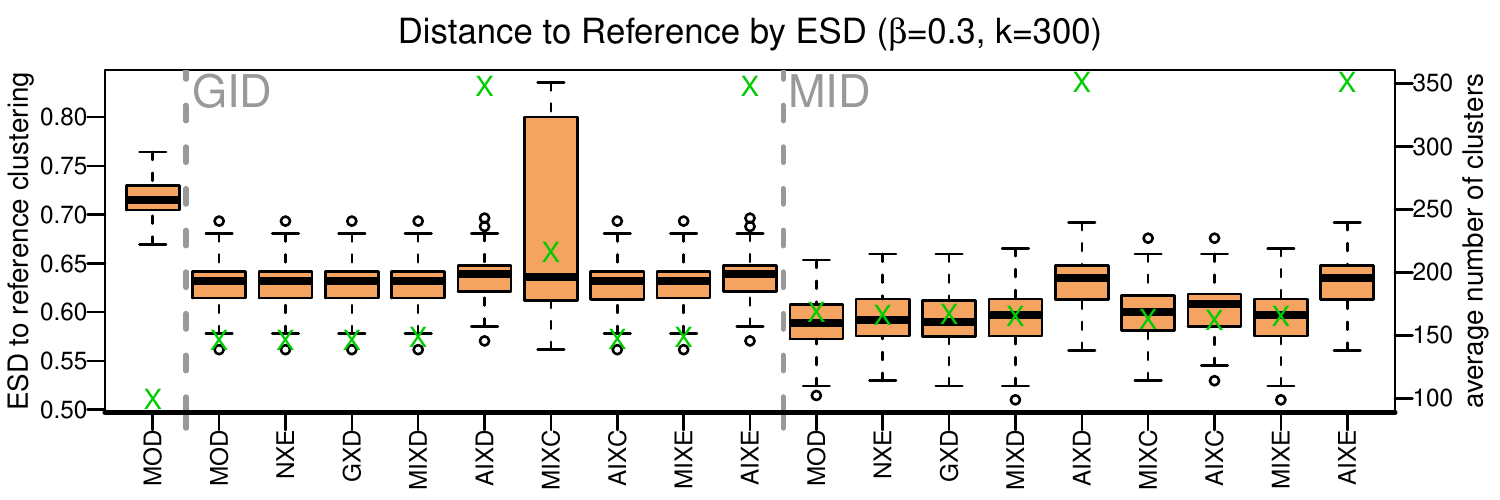}}%
    \end{minipage}%
  }%
\end{figure}
\begin{figure}[t]
  \subfloat{%
    \begin{minipage}[t]{\textwidth}%
      \centerline{\includegraphics[scale=0.8]{distancesToRef_G_ARAND_beta1_k10}}%
    \end{minipage}%
        }%
  \hfill%
  \vspace{0.1cm}
  \subfloat{%
    \begin{minipage}[t]{\textwidth}%
      \centerline{\includegraphics[scale=0.8]{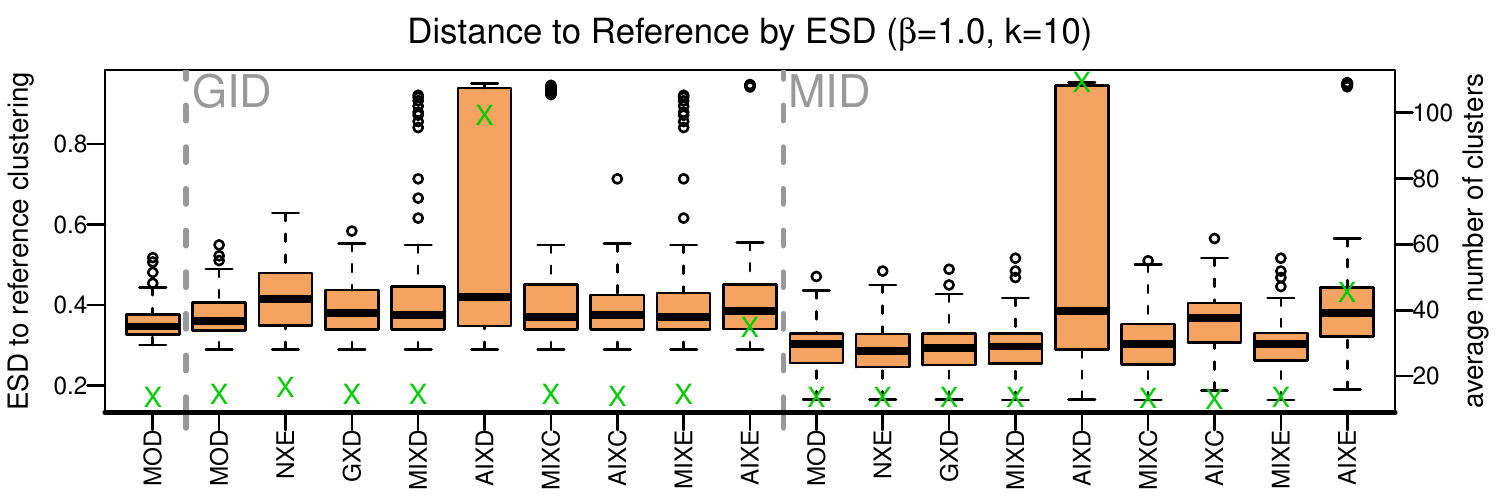}}%
    \end{minipage}%
  }%
  \hfill%
  \vspace{0.1cm}
  \subfloat{%
    \begin{minipage}[t]{\textwidth}%
      \centerline{\includegraphics[scale=0.8]{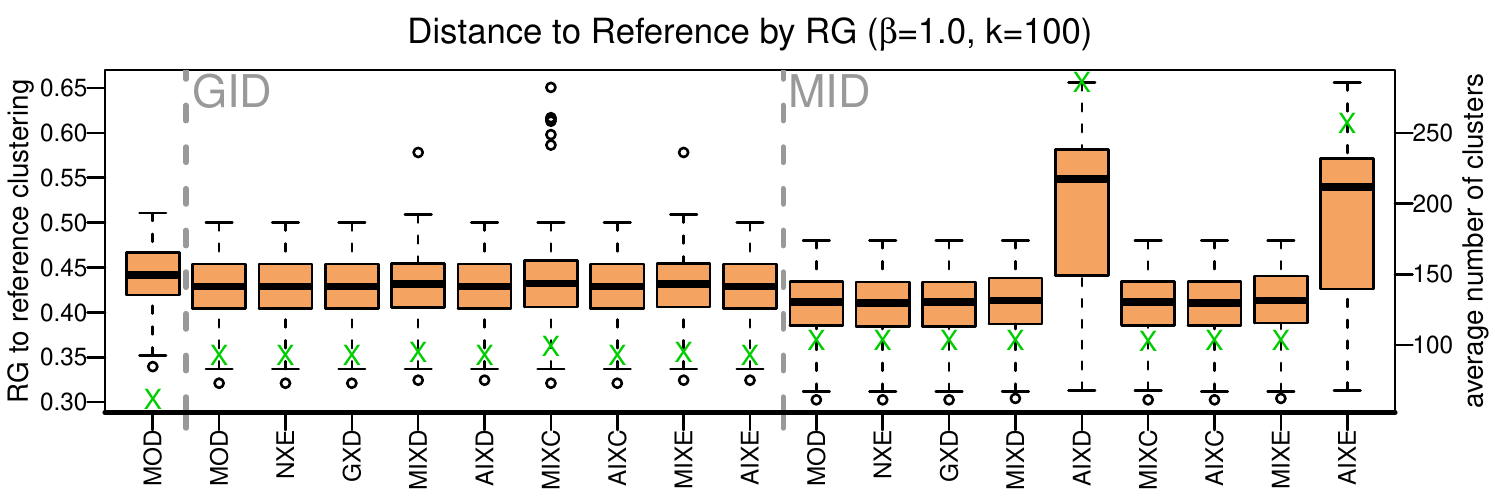}}%
    \end{minipage}%
  }%
    \hfill%
  \vspace{0.1cm}
  \subfloat{%
    \begin{minipage}[t]{\textwidth}%
      \centerline{\includegraphics[scale=0.8]{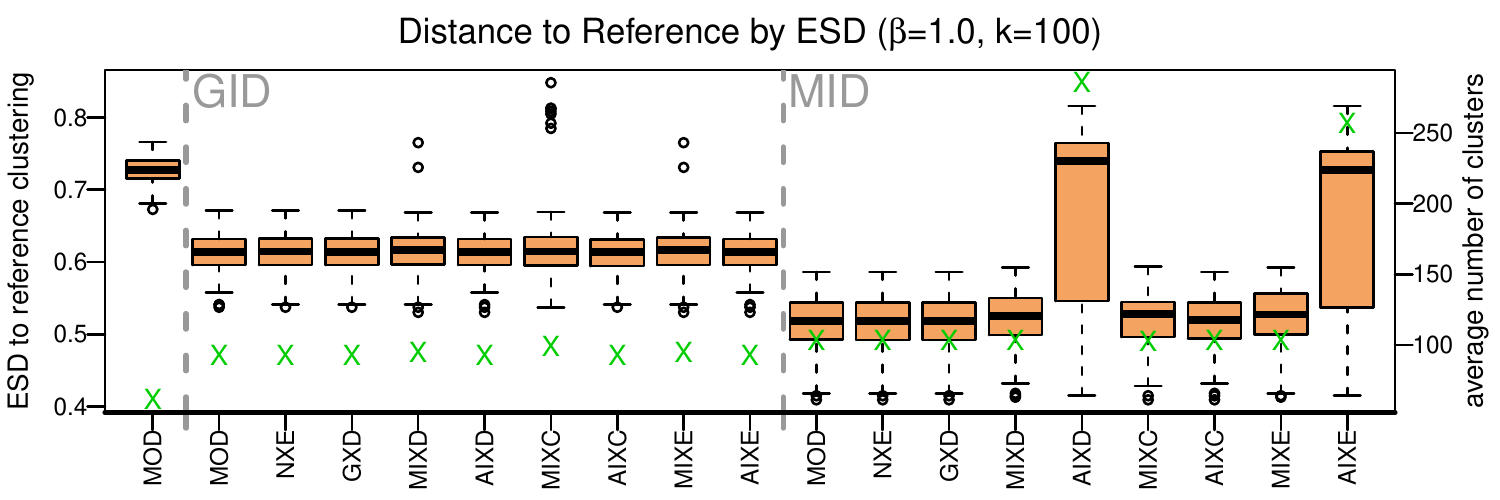}}%
    \end{minipage}%
  }%
\end{figure}
\begin{figure}[t]
  \subfloat{%
    \begin{minipage}[t]{\textwidth}%
      \centerline{\includegraphics[scale=0.8]{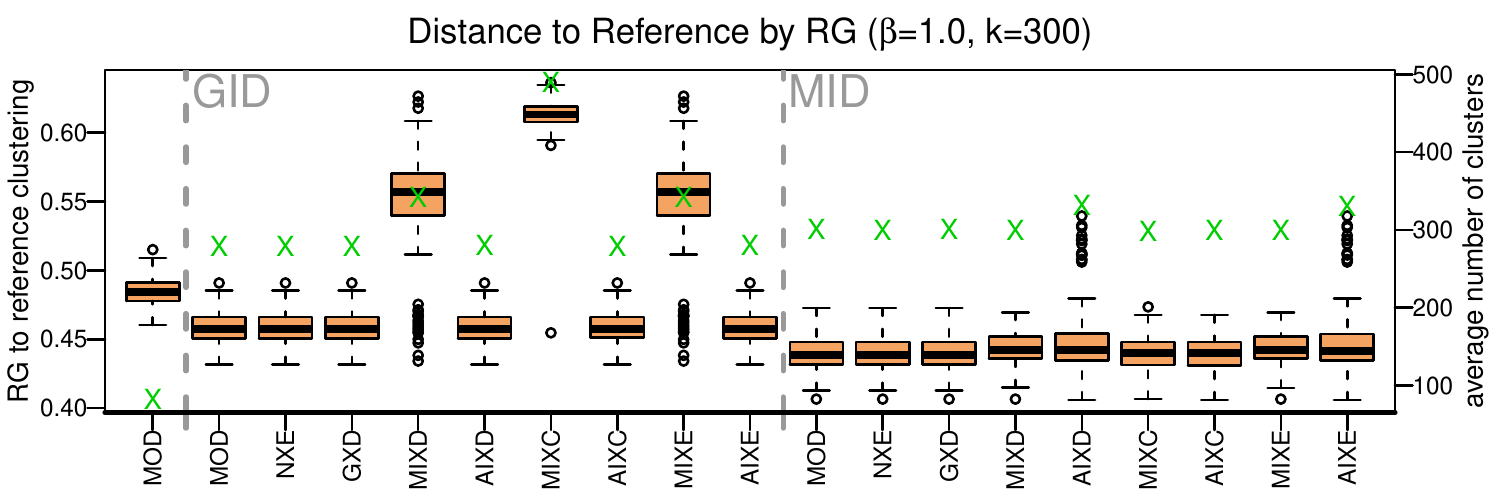}}%
    \end{minipage}%
        }%
  \hfill%
  \vspace{0.1cm}
  \subfloat{%
    \begin{minipage}[t]{\textwidth}%
      \centerline{\includegraphics[scale=0.8]{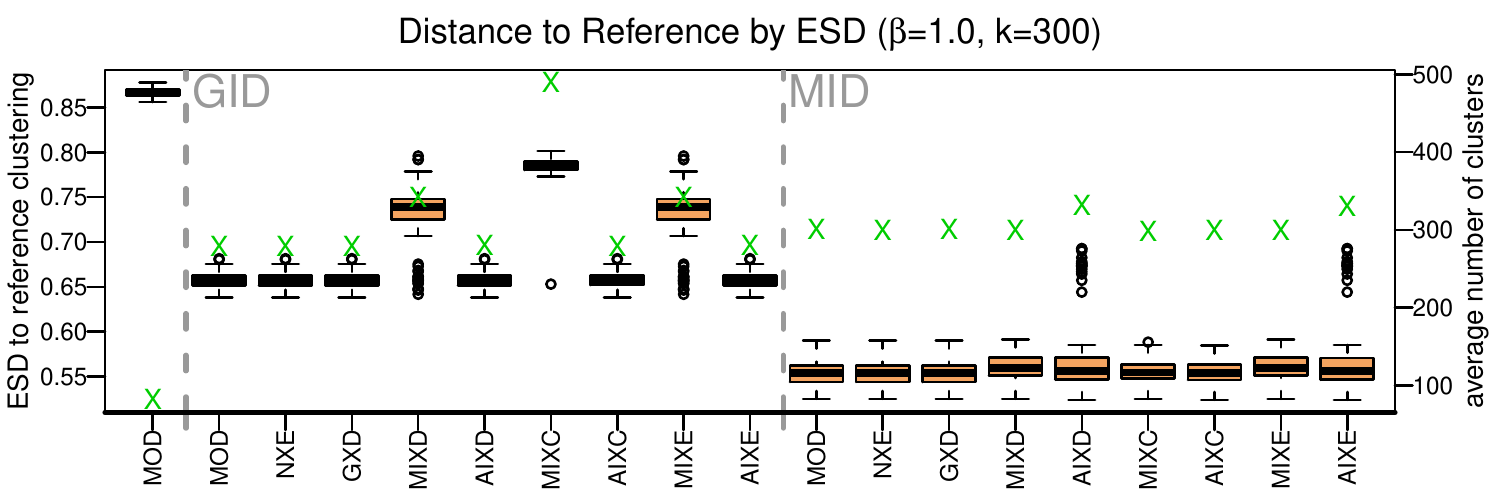}}%
    \end{minipage}%
  }%
  \hfill%
  \vspace{0.1cm}
  \subfloat{%
    \begin{minipage}[t]{\textwidth}%
      \centerline{\includegraphics[scale=0.8]{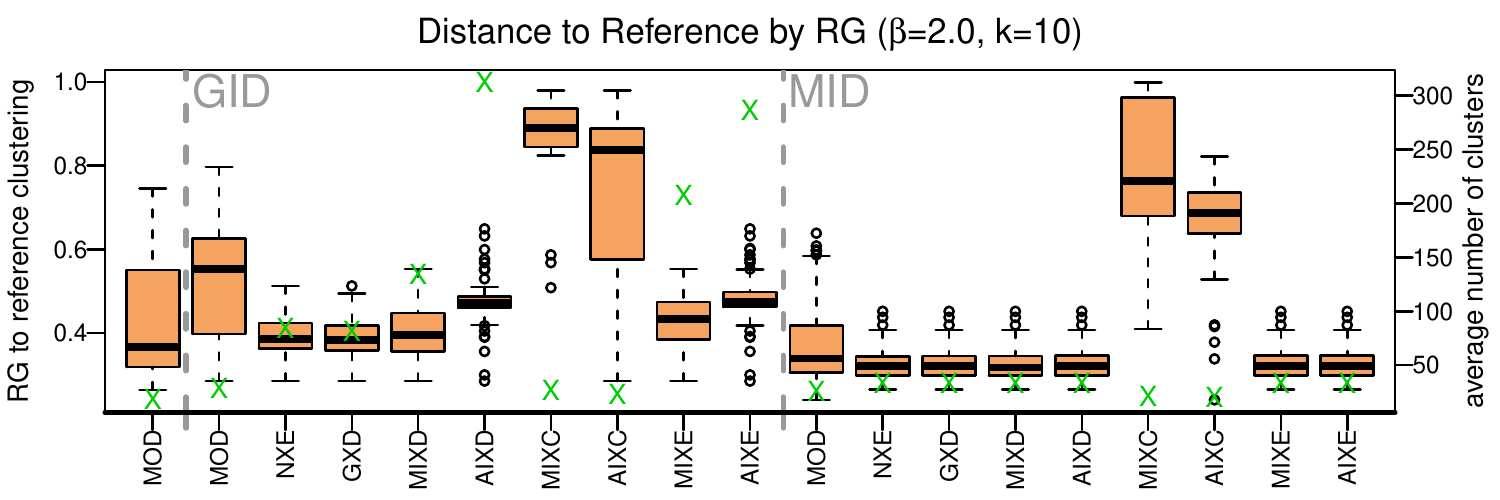}}%
    \end{minipage}%
  }%
    \hfill%
  \vspace{0.1cm}
  \subfloat{%
    \begin{minipage}[t]{\textwidth}%
      \centerline{\includegraphics[scale=0.8]{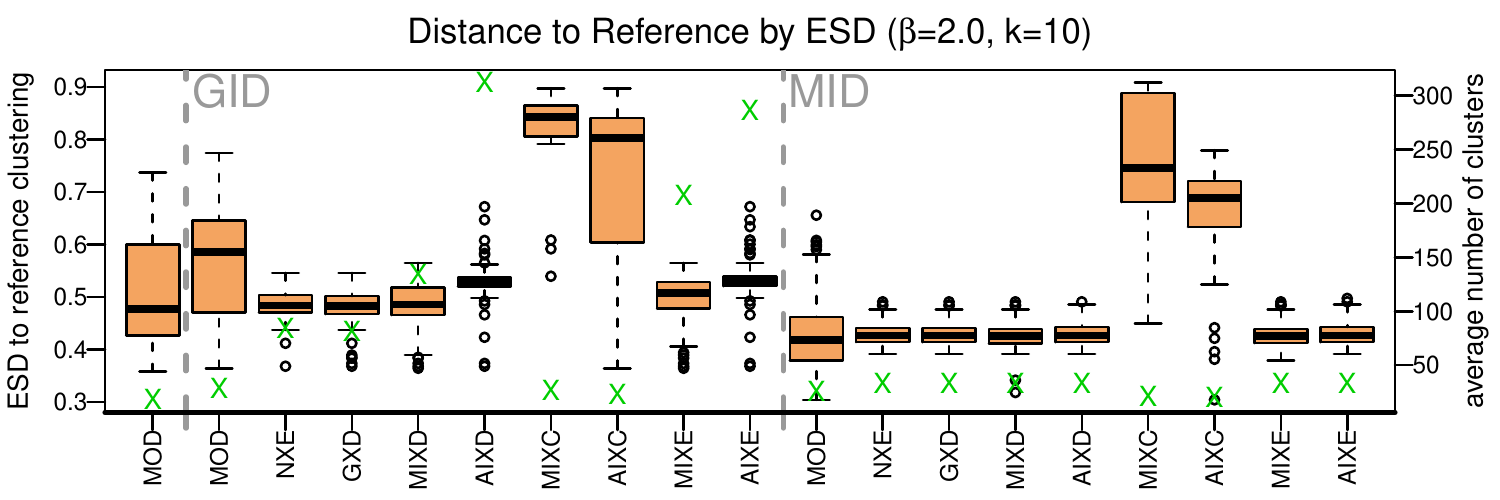}}%
    \end{minipage}%
  }%
\end{figure}
\begin{figure}[t]
  \subfloat{%
    \begin{minipage}[t]{\textwidth}%
      \centerline{\includegraphics[scale=0.8]{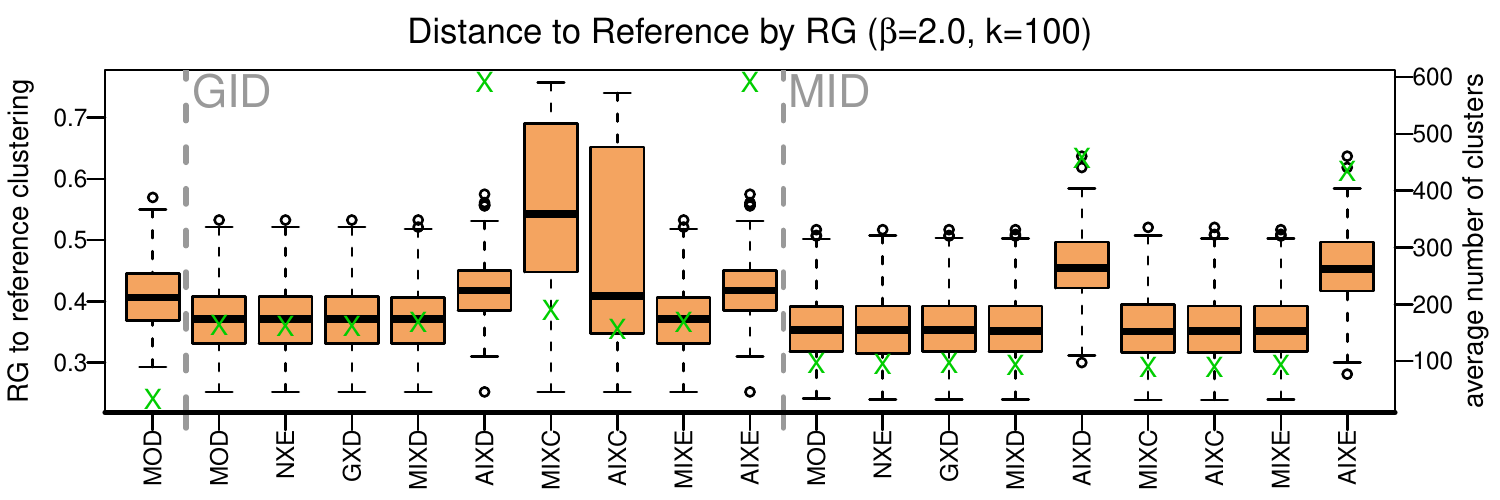}}%
    \end{minipage}%
      }%
  \hfill%
  \vspace{0.1cm}
  \subfloat{%
    \begin{minipage}[t]{\textwidth}%
      \centerline{\includegraphics[scale=0.8]{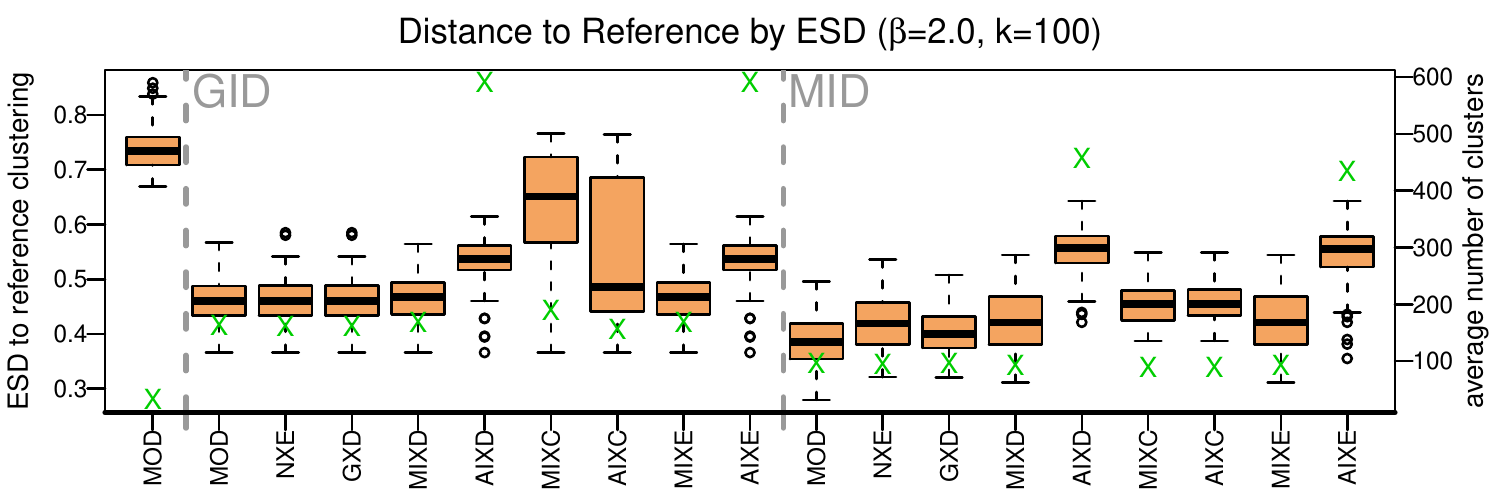}}%
    \end{minipage}%
  }%
  \hfill%
  \vspace{0.1cm}
  \subfloat{%
    \begin{minipage}[t]{\textwidth}%
      \centerline{\includegraphics[scale=0.8]{distancesToRef_G_ARAND_beta2_k300}}%
    \end{minipage}%
  }%
    \hfill%
  \vspace{0.1cm}
  \subfloat{%
    \begin{minipage}[t]{\textwidth}%
      \centerline{\includegraphics[scale=0.8]{distancesToRef_ES_beta2_k300}}%
    \end{minipage}%
  }%
\end{figure}

\end{document}